\documentclass[11pt]{article}
\usepackage{amsthm,amsmath,latexsym,amssymb,amsfonts,amscd}
\usepackage{graphicx}
\usepackage{grffile}
\usepackage[nosort]{cite}

\usepackage{epstopdf}
\usepackage{enumerate}
\usepackage{tensor}
\usepackage{mathrsfs}

\usepackage[english]{babel}
\usepackage{braket}
\usepackage{dcolumn}
\usepackage{bm}
\usepackage{hyperref}
\usepackage{slashed}

\usepackage{comment}

\renewcommand{\theequation}{\thesection.\arabic{equation}}
\makeatletter\@addtoreset{equation}{section}\makeatother

\newcommand{\bep}{\begin{picture}}
\newcommand{\eep}{\end{picture}}

\newcounter{YoungHeight}\newcounter{YoungWidth}

\newcounter{Mul1}\newcounter{Mul2}\newcounter{Mul3}\newcounter{Mul4}
\newcounter{A0}\newcounter{A1}\newcounter{A2}
\newcounter{B3}
\newcounter{C3}\newcounter{C4}
\newcounter{D1}\newcounter{D2}\newcounter{D3}
\newcounter{T0}\newcounter{T1}

\newlength{\txtHShift}

\newlength{\txtWidth}

\newcommand{\HalfLength}[2]{\setcounter{Mul1}{#1}\setcounter{Mul2}{#1}\addtocounter{Mul1}{\value{Mul2}}\addtocounter{Mul1}{\value{Mul2}}%
\addtocounter{Mul1}{\value{Mul2}}\addtocounter{Mul1}{\value{Mul2}}\setcounter{#2}{\value{Mul1}}}

\newcommand{\Add}[3]{\setcounter{#1}{#2}\addtocounter{#1}{#3}}

\newcommand{\Length}[1]{#10}
\newcommand{\YoungScale}{}

\newcommand{\shiftedText}[2]{{\hspace{#1}#2}}
\newcommand{\calcHShift}[1]{\settowidth{\txtWidth}{#1}\setlength{\txtHShift}{-0.5\txtWidth}}

\newcommand{\TextTop}[3]{{\calcHShift{#1}\HalfLength{#2}{T0}\Add{T1}{\Length{#3}}{-9}\put(\value{T0},\value{T1}){\shiftedText{\txtHShift}{#1}}}}

\newcommand{\BlockA}[2]{{\YoungScale\bep(\Length{#1},\Length{#2}){\Add{A1}{#1}{1}\Add{A2}{#2}{1}}%
\multiput(0,0)(10,0){\value{A1}}{\line(0,1){\Length{#2}}}\multiput(0,0)(0,10){\value{A2}}{\line(1,0){\Length{#1}}}%
\setcounter{YoungHeight}{\Length{#2}}\setcounter{YoungWidth}{\Length{#1}}\eep}}

\newcommand{\RectT}[3]{\bep(\Length{#1},\Length{#2})\put(0,0){\line(1,0){\Length{#1}}}\put(0,0){\line(0,1){\Length{#2}}}%
\put(\Length{#1},\Length{#2}){\line(-1,0){\Length{#1}}}\put(\Length{#1},\Length{#2}){\line(0,-1){\Length{#2}}}#3{#1}{#2}\eep}

\newcommand{\RectARow}[2]{{\bep(\Length{#1},10)\put(0,0){\RectT{#1}{1}{\TextTop{#2}}}\eep}}

\newcommand{\YoungA}{\BlockA{1}{1}}

\newcommand{\YoungAA}{\BlockA{1}{2}}

\newcommand{\pl}{\partial}

\newcommand{\be}{\begin{equation}}
\newcommand{\ee}{\end{equation}}

\newcommand{\deh}{\hat{\partial}}

\newcommand{\besubeqs}{\begin{subequations}}
\newcommand{\esubeqs}{\end{subequations}}

\newcommand{\psib}{\bar{\psi}}
\newcommand{\JOOst}[2]{{\overline{\langle J_{#1} O_{#2}O_{#2}\rangle}}}
\newcommand{\Fron}{{\Phi}}
\newcommand{\mm}{{\ensuremath{\underline{m}}}}

\newcommand{\preprint}[1]{\begin{table}[t]    
             \begin{flushright}               
             {#1}                             
             \end{flushright}                 
             \end{table}}                     
\renewcommand{\title}[1]{\vbox{\center\LARGE{#1}}\vspace{5mm}}
\renewcommand{\author}[1]{\vbox{\center#1}\vspace{5mm}}
\newcommand{\address}[1]{\vbox{\center\em#1}}

\begin{document}
\begin{titlepage}
\preprint{PUPT-2517}
\preprint{LMU-ASC 05/17}
\begin{center}
\vskip 1cm

\title{Notes on Spinning Operators in Fermionic CFT}

\author{S. Giombi$^{1}$, V. Kirilin$^{1,4}$, E. Skvortsov$^{2,3}$}

\address{${}^1$Department of Physics, Princeton University, Princeton, NJ 08544}
\address{${}^2$Arnold Sommerfeld Center for Theoretical Physics, Ludwig-Maximilians University Munich, Theresienstr. 37, D-80333 Munich, Germany}
\address{${}^3$Lebedev Institute of Physics, Leninsky ave. 53, 119991 Moscow, Russia}
\address{${}^4$ITEP, B. Cheremushkinskaya 25, Moscow, 117218, Russia}

\end{center}

\abstract{The Gross-Neveu model defines a unitary CFT of interacting fermions in $2<d<4$ which has perturbative descriptions
in the $1/N$ expansion and in the epsilon-expansion near two and four dimensions. In each of these descriptions, the CFT has
an infinite tower of nearly conserved currents of all spins. We determine the structure of the non-conservation equations both
at large $N$ and in the epsilon-expansion, and use it to find the leading order anomalous dimensions of the broken currents. Similarly,
we use the fact that the CFT spectrum includes a nearly free fermion to fix the leading anomalous dimensions of a few scalar composite
operators. We also compute the scaling dimensions of double-trace spinning operators in the large $N$ expansion, which correspond to
interaction energies of two-particle states in the AdS dual higher-spin theory. We first derive these anomalous dimensions
by a direct Feynman diagram calculation, and then show that the result can be exactly reproduced by analytic bootstrap methods, provided the sum
over the tower of weakly broken higher-spin currents is suitably regularized. Finally, we apply the analytic bootstrap approach to derive the
anomalous dimensions of the double-trace spinning operators in the 3d bosonic and fermion vector models coupled to Chern-Simons theory, to
leading order in $1/N$ but exactly in the `t Hooft coupling.}

\vfill

\end{titlepage}

\tableofcontents
\section{Introduction and Summary}
The Gross-Neveu (GN) model \cite{Gross:1974jv}
\begin{equation}
{\cal L}_{\rm GN} = \bar\psi_i \slashed{\partial}\psi^i +\frac{g}{2}(\bar\psi_i\psi^i)^2
\label{GN-lag}
\end{equation}
is a classic example of quantum field theory of interacting fermions. Here $\psi^i$, $i=1,\ldots,N_f$ denotes a collection of $N_f$ Dirac fermions, so
that the theory has a manifest $U(N_f)$ global symmetry. When studied as a function of dimension $d$, there is evidence that the GN model describes
a unitary interacting CFT in $2<d<4$,
which corresponds to a non-trivial UV fixed point of (\ref{GN-lag}). Despite the fact that the quartic interaction
is irrelevant, the model is formally renormalizable in the framework of the $1/N$ expansion \cite{Wilson:1972cf, Gross:1975vu},
and this approach can be used to compute various physical quantities at the interacting fixed point
as a function of $d$ (with $d=3$ being the physically interesting dimension), see \cite{Moshe:2003xn} and references therein for a comprehensive review.
Another approach to the Gross-Neveu CFT is the Wilson-Fisher $\epsilon$-expansion: in $d=2$ the four-fermi interaction is renormalizable,
and working in $d=2+\epsilon$ one finds UV fixed points which are weakly coupled for small $\epsilon$. Critical exponents at these fixed points can be computed
by usual perturbation theory for finite $N_f$. In \cite{ZinnJustin:1991yn, Hasenfratz:1991it}, it was suggested that the fermionic CFT in $2<d<4$ admits yet
another perturbative description near $d=4$, in terms of the Gross-Neveu-Yukawa theory
\begin{align}
{\cal L}_{\rm GNY} =  \frac{1}{2}(\partial_{\mu}\sigma)^{2}+\bar{\psi}_{i}{\not\,}\partial \psi^{i} + g_{1}\sigma \bar{\psi}_{i}\psi^{i}
+\frac{1}{4!}g_{2}\sigma^{4}\ .
\label{GNY-Lag}
\end{align}
Working in $d=4-\epsilon$, one finds stable IR fixed points for any $N_f$, and there is considerable evidence
that these fixed points correspond to the same CFT defined by the UV fixed
point of the GN model. The information from the various perturbative approaches to the fermionic CFT can be used to obtain estimates
for critical exponents and other physical quantities in the physical dimension $d=3$, see for instance
\cite{Karkkainen:1993ef, Fei:2014yja, Diab:2016spb, Fei:2016sgs}.

When the interactions are turned off, the theory of $N_f$ massless fermions defines a unitary CFT in any dimension $d$. Being a free CFT,
it enjoys an exact higher-spin (HS) symmetry and corresponding exactly conserved currents of all spins which are constructed from fermion bilinears.
In general $d$, the spectrum of these currents is more involved than that of the free scalar CFT. There are totally symmetric currents, but also currents
in mixed-symmetry representations of $SO(d)$ that are obtained using the totally antisymmetric products $\gamma_{\nu_1\ldots \nu_k}$ of gamma matrices.
Their explicit construction will be discussed in section \ref{free-fer} below.
In the free CFT, all these currents are conserved; they have exact scaling dimension $\Delta=d-2+s$, where $s$ is the spin, and belong
to short representations of the conformal algebra. When interactions are turned on, the
currents acquire anomalous dimensions and are no-longer exactly conserved
(except for the stress tensor or spin 1 currents corresponding to the global symmetry):
\begin{equation}
\partial \cdot J_s \sim g K_{s-1}\,,
\label{non-conse}
\end{equation}
where $g$ is a parameter that controls the HS symmetry breaking. In the large $N$ expansion we have $g\sim 1/\sqrt{N}$, and in the $\epsilon$-expansion
$g$ is a power of $\epsilon$, so that the HS symmetry is weakly broken at large $N$ or small $\epsilon$.
In the above equation, $K_{s-1}$ denotes an operator of spin $s-1$ and dimension $d-1+s+O(g)$: this is the primary operator of the unbroken
theory ($g=0$) which recombines with $J_s$ to form a long multiplet in the interacting theory. As reviewed in section \ref{gener}, the non-conservation
equation (\ref{non-conse}) can be used to deduce the anomalous dimensions of the broken currents to leading order in $g$, by computing correlators in the unbroken theory
\cite{Anselmi:1998ms, Belitsky:2007jp}.
This method was applied recently in \cite{Skvortsov:2015pea,Giombi:2016hkj} to the scalar $O(N)$ model in its large $N$ and $\epsilon$-expansions, and in
\cite{Giombi:2016zwa} it was also used to extract the $1/N$ anomalous dimensions of HS currents in the bosonic and fermionic 3d Chern-Simons vector models
of \cite{Giombi:2011kc, Aharony:2011jz}. In this paper, we will apply the same method to the critical GN model, both in the large $N$ expansion
for any $d$, and in the $\epsilon$-expansions near $d=2$ and $d=4$, and extract the anomalous dimensions of weakly broken currents to leading order.
At large $N$, we reproduce known results obtained long ago by diagrammatic methods \cite{Muta:1976js}. In $d=2+\epsilon$ and $d=4-\epsilon$, as far as we know,
our results are new. In all cases, we find precise matching of $\epsilon$-expansions and large $N$ (including the recent
$1/N^2$ results of \cite{Manashov:2016uam}) in their overlapping regime of validity, which provides a nice cross-check of the various approaches to
the interacting CFT.

In the context of the AdS/CFT correspondence, the free fermionic CFT$_d$ (restricted to its $U(N_f)$ singlet sector) should be holographically dual
to the so-called ``type B" higher-spin gravity theory in AdS$_{d+1}$, which includes towers of massless higher-spin gauge fields in one-to-one correspondence
with the conserved currents in the boundary CFT$_d$.\footnote{Such type B theory is known at non-linear level only in the case of AdS$_4$ in the form of Vasiliev equations \cite{Vasiliev:1992av},
see \cite{Giombi:2012ms,Didenko:2014dwa,Giombi:2016ejx} for reviews with a focus on AdS/CFT applications. In
general $d$, one can construct the spectrum and free equations of motion of the bulk theory,
and in principle reconstruct interactions order by order in perturbation theory,
but fully non-linear equations of motion of the Vasiliev type \cite{Vasiliev:1992av, Vasiliev:2003ev} are not known.} In this context,
the critical Gross-Neveu CFT can be thought of as a ``double-trace" deformation of the free theory, and it follows from general arguments
\cite{Klebanov:1999tb} that the AdS dual of the UV fixed point should
be the same higher-spin gravity theory, with the choice of alternate boundary condition ($\Delta=1$ instead of $\Delta=d-1$)
on the bulk scalar field dual to the $\bar\psi\psi$ operator, in analogy with the original conjecture \cite{Klebanov:2002ja}
in the case of the $O(N)$ model. With alternate boundary conditions in the bulk, the higher-spin fields are expected to acquire
masses at loop level, corresponding to the fact that the anomalous dimensions start at $1/N$-order on the CFT side. The role of
the Higgs field in the bulk \cite{Girardello:2002pp} is played by a two-particle state with the appropriate quantum numbers, which
should correspond to the operator appearing on the right-hand side of the non-conservation equation (\ref{non-conse}). In the large $N$ CFT
this operator is indeed of the double-trace type, and we will determine its explicit form in section \ref{largeN}. Schematically,
\begin{equation}
\partial \cdot J_s \sim \frac{1}{\sqrt{N}} \sum_{s'<s} \partial^n B_{(s',1)} \partial^m J_0\,,
\label{largeN-schematic}
\end{equation}
where $B_{(s,1)}$ denotes the mixed-symmetry current in the representation $[s,1,0\ldots,0]$ (dual to
the corresponding mixed-symmetry field in the bulk), and $J_0$ denotes the scalar operator with $\Delta=1+O(1/N)$. In particular, this
equation implies that the relevant bulk one-loop diagrams responsible for the anomalous dimensions involve the cubic coupling
of a totally symmetric field, a mixed-symmetry field, and a scalar. It would be interesting to fix the form of these couplings directly
in the bulk. Note that in $d=3$ the mixed-symmetry fields are in fact related to the totally symmetric fields, due to
$\gamma_{\mu\nu} = i\epsilon_{\mu\nu\rho}\gamma_{\rho}$, but this is not so in general $d$, and
it would be interesting to study more generally the 3-point couplings involving mixed-symmetry fields in the bulk.  Note also that
(\ref{largeN-schematic}) contains more information than just the anomalous dimensions of $J_s$: for instance, it implies that the
3-point functions $\langle J_s(x) B_{(s',1)}(y) J_0(z) \rangle$ break the $J_s$ current conservation (for $s>s'$) already at leading order in $N$.

The methods we use to fix the HS anomalous dimensions, based on the idea of multiplet recombination, are closely related to the approach
put forward in \cite{Rychkov:2015naa}, see \cite{Basu:2015gpa,Sen:2015doa,Ghosh:2015opa,Raju:2015fza,Manashov:2015fha,
Bashmakov:2016uqk, Bashmakov:2016pcg, Nii:2016lpa, Roumpedakis:2016qcg, Liendo:2017wsn} for subsequent related work. In this approach, the leading order anomalous dimensions
of various composite operators
in the $\epsilon$-expansion of $O(N)$ or GN models were fixed using conformal symmetry and the required form of multiplet recombination (essentially
dictated by the classical equations of motion) of the nearly free fields $\phi$ or $\psi$.
In section \ref{scalar-comp}, we apply a similar approach to fix the scaling dimensions of some scalar composites in the critical
GN model at large $N$, as well as in the GNY model in $d=4-\epsilon$. We also show how similar methods can be used in the case
of the large $N$ expansion of the scalar $O(N)$ model. In particular, this appears to lead to a relatively simple derivation of the $1/N$ anomalous dimension
of the scalar singlet operator (with $\Delta=1+O(1/N)$ in GN and $\Delta=2+O(1/N)$ in the $O(N)$ model) compared to the traditional diagrammatic
expansion (see e.g. \cite{Vasiliev:1982dc, Vasiliev:1992wr, Gracey:1990wi}).

In section \ref{sigmadsigma}, we move on to study a different type of operators with spin, namely the ``double-trace" operators
$\sim \sigma \partial^s \sigma$ built ouf of the scalar singlet $\sigma \sim \bar\psi\psi$. These operators have twist $2+O(1/N)$ at large $N$,
and for general $d$ they are not almost conserved currents.\footnote{They become conserved in the $d\rightarrow 4$ limit, where they correspond to one of
the two towers of exactly conserved HS operators in the GNY model (\ref{GNY-Lag}) at the $d=4$ trivial fixed point $g_1=g_2=0$. The two
towers non-trivially mix in $d=4-\epsilon$, as explained in section \ref{sec:GNY}.} We compute their anomalous dimensions in section \ref{sigmadsigma}
directly from Feynman diagrams in $1/N$ perturbation theory. The general $d$ result is given in (\ref{fermionsigma}), and in $d=3$ it reads
\begin{equation}
\Delta_{\sigma\partial^s \sigma} -s-2\Delta_{\sigma} =\frac{32}{\pi^2(2s+1)}\frac{1}{N}+O(1/N^2)\,.
\label{3d-sigsig}
\end{equation}
From the AdS point of view, the anomalous dimension defined by the right-hand side has the interpretation of the interaction energy associated to the
two-particle state of two bulk scalar fields with orbital angular momentum $s$. Perhaps surprisingly, we find that this quantity is positive,
corresponding to an effective repulsive interaction, for all spins in $2<d<4$.\footnote{This result is not in violation of Nachtmann's theorem
\cite{Nachtmann:1973mr}, because in $d<4$ the operators $\sim \sigma \partial^s\sigma$ are not the leading twists in the $\sigma\sigma$ OPE, due
to the presence of the nearly conserved HS currents with twist $d-2+O(1/N)$.}
In section \ref{sec:lcone-boots} we compare this result, as well as the
one for the analogous operators (with $\sigma\sim \phi^2$) in the $O(N)$ model \cite{Lang:1992zw}, to the analytic bootstrap analysis
\cite{Fitzpatrick:2012yx, Komargodski:2012ek} (see \cite{Kaviraj:2015cxa, Alday:2015eya,Dey:2016zbg, Kaviraj:2015xsa,Li:2015rfa, Alday:2015ota,Alday:2015ewa,Alday:2016njk, Alday:2016jfr,Simmons-Duffin:2016wlq} for relevant related work)
of the large spin expansion of the anomalous dimensions of double-trace-like operators of the form $O \partial^s O$.
The OPE data needed for the bootstrap analysis is obtained in Appendix \ref{OPEC} from that of the free theories using the AdS/CFT dictionary
for double-trace flows \cite{Klebanov:1999tb,Klebanov:2002ja}. We find
that the $\sigma \partial^s\sigma$ anomalous dimensions in the GN and $O(N)$ model can be exactly reproduced
in the analytic bootstrap approach, provided one suitably regulates the sum over the exchange of the infinite
tower of nearly conserved currents of all even spins.
Even though the contribution of each nearly conserved even spin current to the $\sigma\partial^s\sigma$ anomalous dimension is negative,
the regularized sum over the HS tower appears to yield a final positive result in the GN model in agreement with (\ref{3d-sigsig}),
and a vanishing result for the $d=3$ $O(N)$ model, in agreement with \cite{Lang:1992zw} (see also \cite{Leonhardt:2003du}).
More generally, the arbitrary $d$ results can also be reproduced in the same way.
As a consistency check of the regularized sum over spins, we also show that it correctly implies vanishing of the anomalous dimensions
of the double-trace operators in the free fermionic and scalar CFT in any $d$. Finally, in section \ref{CSmat} we use the same analytic
bootstrap approach to compute the anomalous dimensions of the same type of double-trace operators in the bosonic and fermionic vector models
coupled to Chern-Simons gauge theory in $d=3$ \cite{Giombi:2011kc,Aharony:2011jz}, working to leading
order in $1/N$ but exactly in the `t Hooft coupling $\lambda$. In the CS-fermion model, the anomalous dimensions vanish to the order $1/N$ for all $\lambda$,
and in the CS-scalar model they are given by an expression that smoothly interpolates between the free scalar at $\lambda=0$ and the critical GN model
at $\lambda\rightarrow 1$, in agreement with the conjectured 3d bosonization duality \cite{Maldacena:2012sf,Aharony:2012nh}.

\section{Free Fermions}
\label{free-fer}
Let us consider the free CFT of $N_f$ massless Dirac fermions. For general $d$,
the spectrum of bilinear primary operators is more complicated than that of the free scalar CFT.
In addition to a tower of totally-symmetric conserved tensors $J_{\mu_1 \cdots \mu_s}$, as in the free scalar theory, and the scalar operator $J_0=\bar\psi\psi$ of dimension $\Delta=d-1$, we have towers of conserved tensors of mixed-symmetry $B_{\mu_1\cdots \mu_s,\nu_1\cdots \nu_k}$
and a finite number of anti-symmetric tensors $B_{\nu_1\cdots \nu_k}$ that are not conserved currents, see e.g. \cite{Vasiliev:2004cm,Alkalaev:2012rg}.

\subsection{Totally symmetric higher-spin currents}
It is convenient to introduce an auxiliary null vector  $z^{\mu}$ in order to contract the indices of a symmetric traceless tensor
\begin{align}
    \hat{J}_{s}(x,z)=J_{\mu_1...\mu_s}(x)z^{\mu_1}...\,z^{\mu_s}\,.
\end{align}
One may restore the explicit indices on the currents by acting with the differential operator in $z$-space (sometimes called Thomas derivative)
\cite{Dobrev:1975ru, Craigie:1983fb, Belitsky:2007jp, Costa:2011mg}:
\be
\label{Dz}
D^{\mu}_z\equiv \left(\frac{d}{2}-1\right)\partial_{z_\mu} +z^\nu \partial_{z_\nu} \partial_{z_\mu} - \frac{1}{2} z^\mu \partial_{z_\nu} \partial_{z_\nu}.
\ee
The explicit form of the currents can be conveniently given as
\begin{equation}
\label{fermcurgen}
 (\hat{J}_{s})^i_{\ j}=f_s(\deh_1, \deh_2) \bar{\psi}_j (x_1) \hat{\gamma} \psi^{i}(x_2)\Big|_{x_{1,2}=x}\,,\qquad s\ge 1\,,
\end{equation}
where $\deh_{1,2} \equiv z\cdot \partial_{1,2}$, and $f_s(u,v)$ is a homogeneous function of total degree $s-1$. Here $i,j=1,\ldots,N_f$
are the flavor indices, and we can of course decompose $(J_s)^{i}_{\ j}$ into the $U(N_f)$ singlet part, and the adjoint (traceless)
currents $(J_s^{A})^i_{\ j} \sim \bar\psi_j \deh^{s-1} \hat\gamma \psi^i-\frac{1}{N_f}\delta^i_j\bar\psi \deh^{s-1}\hat\gamma \psi$. In
the remaining of this section we will mostly omit flavor indices for simplicity.

Imposing the conservation condition  $\partial_{\mu} D^{\mu}_z \hat{J}_s=0$ and using the free Dirac equation
one finds for $f_s(u,v)$:
\begin{align}
\label{f-diffeq}
  \big(\frac{d}{2} (\partial_u + \partial_v) + u\partial^2_u + v\partial^2_v \big) f_s = 0\,.
\end{align}
The solution is given by
\begin{align}
    f_s =(\deh_1+\deh_2)^{s-1} C^{d/2-1/2}_{s-1} \Big(\frac{\deh_1-\deh_2}{\deh_1+\deh_2} \Big)\,,
\end{align}
which takes the same form as the free scalar CFT (see e.g. \cite{Skvortsov:2015pea,Giombi:2016hkj}), up to the shifts $s \rightarrow s-1$, $d \rightarrow d+2$.
Alternatively, we can obtain the same differential equation (\ref{f-diffeq}) by imposing that (\ref{fermcurgen})
is a conformal primary (see e.g. \cite{Craigie:1983fb}). Of course, these operators have exact dimension $\Delta_s =d-2+s$ in the free CFT.

\subsection{Mixed-symmetry currents}
The currents constructed above are totally symmetric, corresponding to the representation $(s,0,0,\ldots)$ of $SO(d)$.
In general dimension $d$, there also exist conserved tensor primaries of the symmetry $(s,1,..,1,0,\ldots)$, corresponding to the Young diagram
\begin{align}
\parbox{60pt}{\bep(60,60)\put(0,40){\YoungA}\put(0,50){\RectARow{5}{$s$}}\put(0,0){\YoungAA}\put(3,25){\vdots}\eep}
\end{align}
All such mixed-symmetry currents can in principle be extracted from a simple generating formula \cite{Alkalaev:2012rg}
\begin{align}\label{redgenfunc}
\tilde{B}_{\nu_1...\nu_k} (x) &=\bar{\psi}(x+y)\gamma_{\nu_1...\nu_k}\psi(x-y) \Big|_{y=0}\,,
\end{align}
where $\gamma_{\nu_1...\nu_k}\equiv \gamma_{[\nu_1}...\gamma_{\nu_k]}$ is the totally anti-symmetrized product of $\gamma$-matrices.
One can easily show that it generates conserved currents except for the case of the totally anti-symmetric primaries $(1,...,1)$
\begin{align}
\bar{\psi} \gamma_{\nu_1...\nu_k}\psi\,,& && k>1\,,
\end{align}
which are not conserved and should be AdS/CFT dual to anti-symmetric massive fields.
However, the simple generating function \eqref{redgenfunc}, when expanded in $y$,
does not give conformal primaries, but a mixture with descendants (the expansion of (\ref{redgenfunc}) does not produce irreducible
tensors). To obtain the primary operators, let us look for the generating function
\begin{align}
{B}_{\nu_1...\nu_k}&=\sum_s \frac{1}{s!} B_{\mu_1...\mu_s,\nu_1...\nu_k}z^{\mu_1} \cdots z^{\mu_s}
\end{align}
where, as for the totally symmetric tensors, we use a null polarization vector to contract all the symmetric indices.
The mixed-symmetry primaries have to obey a number of irreducibility conditions:
\begin{align}
B_{(\mu_1...\mu_s,\mu_{s+1})\nu_1...\nu_k}&=0\,, && \delta^{\rho\sigma} B_{\mu_1...\mu_{s-2}\rho\sigma,\nu_1...\nu_k}=0\,,\\
\pl^\lambda B_{\mu_1...\mu_{s-2}\lambda[\nu_0,\nu_1...\nu_k]}&=0\,.
\end{align}
Here symmetrization over all $\mu$ indices and anti-symmetrization over all $\nu$ indices is implied, which is indicated by the brackets. The first condition imposes $(s,1,\ldots,1,0, \ldots )$
symmetry; the second one tells that the tensor is traceless in all the indices provided the first condition is satisfied; the third one implies that the divergence projected onto the $(s-1,1\ldots,1,0,\ldots)$ symmetry vanishes (there are two independent divergences: $(s,1,\ldots,1,0,\ldots)$ and
$(s-1,1,\ldots,1,0,\ldots)$ and only the latter is the primary that needs to be decoupled).
The most general ansatz for the generating function reads:\footnote{In principle, one can introduce auxiliary anti-commuting variables as to hide the $\nu$ indices and work out the super-symmetric Thomas derivative. Fortunately we will need only the simplest mixed-symmetry currents.}
\begin{align}\label{mscurrent}
\begin{aligned}
&{B}_{\nu_1...\nu_k}(\hat{\pl}_i; z)=F_1\bar{\psi}(x_1)\gamma_{\nu_1...\nu_k\rho}z^\rho\psi(x_2)+F_2\bar{\psi}(x_1)\gamma_{[\nu_1...\nu_{k-1}}z_{\nu_k]}\psi(x_2)+\\&\qquad+F_3 \pl_{[\nu_1} \bar{\psi}(x_1)\gamma_{\nu_2...\nu_{k-1}\rho}z^\rho z_{\nu_k]}\psi(x_2)+F_4  \bar{\psi}(x_1)\gamma_{[\nu_1...\nu_{k-2}\rho}z^\rho z_{\nu_{k-1}}\pl_{\nu_k]}\psi(x_2)\,,
\end{aligned}
\end{align}
where again anti-symmetrization over $\nu$ indices is implied. Functions $F_{1,2,3,4}$ depend on $\hat{\pl}_{1,2}$. The usage of the null polarization vector $z^\mu$ takes away the traces in the $\mu$ indices. However, the trace with respect to one symmetric and one antisymmetric index $\delta^{\mu\nu}$ needs to be subtracted by hand. Altogether, the Young, the conservation and the tracelessness conditions, when expressed in terms of the generating function, give:
\begin{align}
z^\rho B_{\rho\nu_1...\nu_{k-1}}&=0\,, &
 \partial_{\mu} D^{\mu}_z D^z_{[\nu_1}B_{\nu_2...\nu_{k+1}]}&=0\,, &
D_z^\mu B_{\mu\nu_1...\nu_{k-1}}&=0\,,
\end{align}
where in the second expression the anti-symmetrization over all $\nu$'s is implied. The trace with respect to $z$ and a free index $\mu$ has to be taken with the help of the Thomas derivative \eqref{Dz}.

In the following we would like to compute the anomalous dimensions of the totally-symmetric higher-spin currents.
The non-conservation operator of those, as will be shown below, contains no more than two gamma-matrices. Therefore it will only
involve the simplest mixed-symmetry primaries with symmetry of the hook diagram
\begin{align}
\parbox{60pt}{\bep(60,20)\put(0,00){\YoungA}\put(0,10){\RectARow{5}{$s$}}\eep}\,.
\end{align}
A simplification occurs in this case and only two terms of \eqref{mscurrent} survive
\begin{align}
\label{mixed}
B_\mu&=F_1\bar{\psi}(x_1)\gamma_{\mu
\nu}z^\nu\psi(x_2)+F_2\bar{\psi}(x_1)z_\mu\psi(x_2)\,.
\end{align}
The Young condition is trivial here and the conservation/tracelessness can be read from
\begin{align}
 \partial_{\mu} D^{\mu}_z D_{z [\nu_1} B_{\nu_2]}&=0\,, & D_z^\mu B_\mu&=0\,.
\end{align}
Solving these equations, we find that
the result for the $(s,1)$ mixed-symmetry currents is
\begin{align}
B_\mu(x,z)&= F_1\bar{\psi}(x_1)\gamma_{\mu\nu}z^\nu\psi(x_2)+F_2\bar{\psi}(x_1)z_\mu\psi(x_2) \Big|_{x_{1,2}=x}\,,\\
&F_1=(\deh_1+\deh_2)^{s-1} C^{d/2-1/2}_{s-1}(w)\,,\qquad\qquad F_2=(\deh_1+\deh_2)^{s-1} C^{d/2-1/2}_{s-2}(w)\,,
\end{align}
where $w={(\deh_1-\deh_2)}/({\deh_1+\deh_2})$, and it is understood that $x_{1,2}\rightarrow x$ after taking all derivatives. Let us give few examples. The $s=1$ case is trivial --- it is not a current:
\begin{align}
B_{\mu}&= \bar{\psi}\gamma_{\mu\nu}\psi z^\nu\,.
\end{align}
The simplest genuine mixed-symmetry current is $(2,1)$ (see also \cite{Alkalaev:2012rg} for the index form):
\begin{align}
B_\mu= (d-1)(\deh_1-\deh_2) \bar{\psi}\gamma_{\mu\nu}z^\nu\psi+(\deh_1+\deh_2) z_\mu \bar{\psi}\psi\,.
\end{align}

Note that while the divergence of the mixed-symmetry current $\partial^{\mu}B_{\mu\mu_2,\ldots \mu_{s-1}[\nu_0,\nu_1]}$ that has $(s-1,1)$ symmetry does vanish, but the divergence with respect to the $\nu$ index is not zero. It defines a descendant
\begin{align}\label{divcurrent}
    \pl^\nu B_\nu &= F_d(\deh_1,\deh_2) \bar{\psi}\psi\,,\\
    F_d(u,v)&=(v-u) (u+v)^{s-1} C_{s-1}^{d/2-1/2}\left(\frac{u-v}{u+v}\right)+(u+v)^s C_{s-2}^{d/2-1/2}\left(\frac{u-v}{u+v}\right)\,,
\end{align}
which will be shown below to naturally enter the non-conservation operator of totally symmetric currents in the interacting CFT.

\subsection{Two-point functions}
The two-point functions of the totally-symmetric currents can be computed as in \cite{Skvortsov:2015pea,Giombi:2016hkj} by using the Schwinger representation for the two-point function $\langle\psi\bar{\psi}\rangle$, which in our conventions reads
\begin{align}
\langle \psi^i (x_1) \bar{\psi}_j (x_2) \rangle = \delta^{i}_j\frac{C_{\psi\psi}\slashed{x}_{12}}{(x^2_{12})^{\Delta_\psi+\tfrac12}}\,, &&  C_{\psi \psi} = \frac{\Gamma(\frac{d}{2})}{2\pi^{d/2} }\,,
\end{align}
so that we can write
\begin{align}
\label{schw}
  \langle \psi (x) \bar{\psi} (0) \rangle   =  -\frac{\Gamma\left(d/2-1\right)}{4\pi^{d/2}}\slashed{\partial} \frac{1}{(x^2)^{d/2-1}} = - \slashed{\partial} \int^{\infty}_{0} \frac{d \alpha}{4\pi^{d/2}} \alpha^{d/2-2} e^{-\alpha x^2}\,.
\end{align}
Using this, one finds after integration over the Schwinger parameters
\begin{align}
\langle J_s(x,z) J_s(0,z)\rangle&= C_{ss} \times \frac{(z\cdot x)^{2s}}{(x^2)^{d+2s-2}}\,, \\
C_{ss}&=N\frac{\pi 2^{-2 d+2 s+1} \Gamma (d+s-2) \Gamma (d+2 s-3)}{\pi^d \Gamma \left(\frac{d-1}{2}\right)^2 \Gamma (s)}\label{Cs}\,,
\end{align}
where $N=N_f \text{tr} \boldsymbol{1}$ is the total number of fermion components. The non-singlet currents have $\text{tr} \boldsymbol{1}$ instead of $N$.

Also, we will need the two-point functions of the hook currents and their descendants \eqref{divcurrent}, which are given by a two-by-two matrix:
\begin{align}
&\begin{pmatrix}
\langle B^s_{\nu}\eta^\nu B^s_\mu\eta^\mu\rangle & \langle \pl^\nu B^s_{\nu} B^s_\mu\eta^\mu\rangle\\
\langle  B^s_{\nu}\eta^\nu  \pl^\mu B^s_\mu\rangle & \langle \pl^\nu B^s_{\nu} \pl^\mu B^s_\mu\rangle
\end{pmatrix}
= N C_s \times  \frac{(z\cdot x)^{2s-2}}{(x^2)^{d+2s-2}}\times\\
& \times \begin{pmatrix}
\eta^2 (z\cdot x)^2 -2(z\cdot x)(\eta\cdot x)(z\cdot \eta)+\frac{(d+2s-4)}{2(d+s-3)}x^2(z\cdot \eta)^2   &  (z\cdot x) (z\cdot \eta) \frac{s(d-2)}{d+s-3}\\
(z\cdot x) (z\cdot \eta)  \frac{-s(d-2)}{d+s-3} &\frac{2 (d-2) s (d+2 s-2)}{d+s-3} \frac{(z\cdot x)^2}{x^2}
\end{pmatrix}\notag\,,
\end{align}
where we introduced an additional vector $\eta^\mu$ to hide the index $\nu$ away. The overall factor is the $C_s$, \eqref{Cs}, from the two-point function of the symmetric currents.

\subsection{Some OPE coefficients}
It is in principle straightforward to work out 3-point (or higher) correlation functions by similar methods. As an example, the 3-point function
of the totally symmetric currents and two fermions is found to be (omitting flavor indices for simplicity, and denoting by $\alpha,\beta$ the spinor
indices)
\begin{equation}
\begin{aligned}
&\langle J_s(x_1,z) \psi^{\alpha} (x_2)\bar\psi_{\beta} (x_3)\rangle = C_{\psi\psi}^2 C_{s\psi\psi}
\left(\slashed{x}_{12}z\cdot \gamma \,\slashed{x}_{13}\right)^{\alpha}_{\ \beta}
\frac{\left(\frac{z\cdot x_{12}}{x_{12}^2}-\frac{z\cdot x_{13}}{x_{13}^2}\right)^{s-1}}{x_{12}^d x_{13}^d}\,,\\
&C_{s\psi\psi} = \frac{(-1)^s 2^{s-1}\Gamma \left(\frac{d}{2}+s-1\right) \Gamma (d+s-2)}{\Gamma (d-1) \Gamma \left(\frac{d}{2}\right) \Gamma (s)} \,,
\end{aligned}
\end{equation}
and similarly one may derive the 3-point functions with mixed-symmetry operators.
In the following, we will also need the 3-point function of $J_s$ with two $\Delta=d-1$ scalar bilinears. A short calculation using the Schwinger
representation and the generating function for $J_s$ yields
\begin{equation}
\begin{aligned}
&\langle J_s(x_1,z) \bar\psi \psi  (x_2)\bar\psi \psi (x_3)\rangle = C_{s00} \frac{\left(\frac{z_1\cdot x_{13}}{x^2_{13}}-\frac{z_1\cdot x_{12}}{x^2_{12}}\right)^s}{x^{d-2}_{12}x^{d-2}_{13}x^{d}_{23}}\,,\\
&C_{s00}=2^{s-1} N C_{\psi\psi}^3(1+(-1)^s) \frac{\Gamma\left(s+\frac{d}{2}-1\right)}{\Gamma\left(\frac{d}{2}\right)}
\frac{\Gamma\left(d+s-2\right)}{\Gamma\left(d-1\right) \Gamma\left(s\right)}\,.
\label{Cs00-fer}
\end{aligned}
\end{equation}
It is instructive to compare this result with the conformal block expansion of the 4-point function of the $\bar\psi\psi$ operator. An explicit calculation yields
\begin{equation}
\begin{aligned}
&\langle \bar\psi\psi(x_1)\bar\psi\psi(x_2)\bar\psi\psi(x_3)\bar\psi\psi(x_4)\rangle = N^2C_{\psi\psi}^4 \frac{g(u,v)}{(x_{12}^2x_{34}^2)^{d-1}}\,, \qquad u = \frac{x_{12}^2 x_{34}^2}{x_{13}^2 x_{24}^2}\,, \qquad v=\frac{x_{14}^2 x_{23}^2}{x_{13}^2 x_{24}^2}\,,\\
&g(u,v) = 1+u^{d-1} +\left(\frac{u}{v}\right)^{d-1}+\frac{1}{N}\left( \frac{u}{v} \right)^{d/2} \left( u^{\frac{d}{2}} - u^{\frac{d}{2}-1}(1+v) +\frac{(1-v)(1-v^{\frac{d}{2}})}{u}- (1+v^{\frac{d}{2}})\right)
\end{aligned}
\label{4pt}
\end{equation}
The function $g(u,v)$ has the conformal block expansion $g(u,v)=1+\sum_{\tau,\ell} a_{\tau,\ell} g_{\tau,\ell}(u,v)$, with $\tau=\Delta-\ell$ the twist of the intermediate state, and $a_{\tau,\ell}$ are related to squares of the OPE coefficients. In the limit $u\rightarrow 0$, $g_{\tau,\ell}(u,v)$ reduces to the so-called collinear conformal blocks
\begin{equation}
g_{\tau,\ell}(u,v) \simeq  u^{\tau/2} (-\tfrac{1}{2})^{\ell}(1-v)^{\ell} {}_2 F_1(\frac{\tau}{2}+\ell,\frac{\tau}{2}+\ell,\tau+2\ell ; 1-v)\,.
\end{equation}
The term of order $u^{(d-2)/2}$ in the small $u$ expansion of (\ref{4pt}) should be reproduced by the sum over the exchanged conserved currents $J_s$ of all even spins, with $\tau=d-2$. Using the OPE coefficients in $C_{s00}$, and $a_s=C_{s00}^2/(C_{ss} N^2 C_{\psi\psi}^4)$ (see the Appendix), we have verified that indeed
\begin{equation}
\sum_{\ell} a_{\ell}  (-\tfrac{1}{2})^{\ell}(1-v)^{\ell} {}_2 F_1(\frac{d-2}{2}+\ell,\frac{d-2}{2}+\ell,d-2+2\ell ;1-v) = \frac{1}{N}\frac{1}{v^{d/2}}(1-v)(1-v^{d/2})\,.
\end{equation}


\section{Weakly broken currents in fermionic CFT}

\subsection{Generalities}
\label{gener}
In this section we review the derivation of the formula which relates the anomalous dimensions of operators in the conformal field theory to the two-point function of the corresponding non-conservation operators \cite{Anselmi:1998ms, Belitsky:2007jp}. Let us begin by noticing that in an arbitrary interacting theory, the operator $\pl_\mu D^\mu_z$ no longer annihilates the  currents (\ref{fermcurgen}), but instead defines the non-conservation operator:
\be
\pl_\mu D^\mu_z \hat{J}_s =  \hat{K}_{s-1}\,.
\ee
Now recall that the two-point function a spin-$s$ primary operator
of dimension $\Delta_s$ is
fixed by conformal invariance to be
\be
\label{conftwopoint}
\langle \hat{J}_s (x_1,z_1) \hat{J}_{s}(x_2,z_2)\rangle =
C_{ss} \frac{\left(z_1\cdot z_2-\frac{2z_1\cdot x_{12} z_2\cdot x_{12}}{x_{12}^2}\right)^{s}}{(x_{12}^2)^{\Delta_{s}}}\,,
\ee
where $z_1$, $z_2$ are two polarization vectors. Writing $\Delta_s = d-2+s+\gamma_s$
and taking the divergence on $x_1$ and $x_2$ on both sides of this equation and setting $z_1=z_2$,
one may derive the following formula for the anomalous dimension, valid to leading order in the breaking parameter
\begin{equation}
\label{mainform}
\gamma_s = -\frac{1}{s (s+d/2 -2) (s+d/2-1)(s+d-3)}
 \frac{(z\cdot x)^2\langle {\hat K}_{s-1} (x,z){\hat K}_{s-1}(0,z)\rangle_0}{\langle \hat{J}_s (x,z) \hat{J}_s(0,z)\rangle_0}\,,
\end{equation}
where the subscript `0' means that the correlators are computed in the ``unbroken" theory. Although for simplicity we have omitted flavor indices,
this formula applies in the same way for singlet and non-singlet currents. In the following we will denote by $\gamma_s$ the anomalous dimension
of the singlet currents, and $\gamma_s^A$ the one of the non-singlets (adjoint).

 To derive the explicit formula for the non-conservation in the various models we consider below,
we act with an operator $\pl_\mu D^\mu_z$ on the currents (\ref{fermcurgen}), which gives terms proportional to the ``descendant operators" $\pl^{\mu}\bar{\psi}\gamma_{\mu}$, $\slashed{\pl} \psi$ and $\partial^2{\bar{\psi}}$, $\partial^2{\psi}$, which are non-zero in the interacting fermion theory:
\begin{eqnarray}
\label{nonconsgen}
&& \pl_{\mu} D_z^{\mu} f_{s}(\deh_1,\deh_2)\bar{\psi}(x_1)\hat{\gamma}\psi(x_2)
= \\
&&\qquad\qquad=\left[\slashed{\pl}_1 q_s(\deh_1,\deh_2) + \slashed{\pl}_2  \tilde{q_s}(\deh_1,\deh_2) +\hat{\gamma} \partial^2_1   h_s(\deh_1,\deh_2)
+ \hat{\gamma} \partial^2_2  \tilde{h}_s(\deh_1,\deh_2) \right]\bar{\psi}(x_1)\psi(x_2)\,,  \cr
&& q_s(u,v)=\big( (\tfrac{d}{2}-1) f_{s}+ v (\partial_v f_{s} - \partial_u f_{s}) \big) \,,\qquad
\tilde q_s(u,v)= \big( (\tfrac{d}{2}-1) f_{s} + u (\partial_u f_{s} - \partial_v f_{s}) \big)\,,\cr
&&h_s(u,v)= \big(\tfrac{d}{2} \partial_u f_{s} + \frac{u-v}{2} \partial^2_u f_{s} + v \partial_{u v} f_{s} \big)\,,\\
&&\tilde h_s(u,v)=\big(t\tfrac{d}{2} \partial_v f_{s} + \frac{v-u}{2} \partial^2_v f_{s} + u \partial_{u v} f_{s} \big)\nonumber
\end{eqnarray}
Explicitly carrying out the differentiation and using recurrence relations for Gegenbauer polynomials we may represent the functions introduced above as:
\begin{eqnarray}
&&q_s(u,v) \equiv q_s^d(u,v) = (u+v)^{s-1} \big[(\tfrac{d}{2} - 1) C^{d/2-1/2}_{s-1}\Big(\frac{u-v}{u+v}\Big) -\frac{2(d-1)v}{u+v}C^{d/2+1/2}_{s-2}\Big(\frac{u-v}{u+v}\Big)\big] ,  \cr
&&\tilde q_s(u,v)=(-1)^{s-1} q^d_s(v,u), \cr &&h_s(u,v) =  (d-1)q^{d+2}_{s-1} (u,v)\,, \\ &&\tilde h_s(u,v) = (d-1)(-1)^{s-1} q^{d+2}_{s-1} (v,u)\,. \nonumber
\end{eqnarray}

\subsection{Large $N$ expansion}
\label{largeN}
One begins with the action of the Gross-Neveu model
\begin{align}
S&=\int d^dx\, \left(\bar{\psi}\slashed{\pl}\psi +\frac12 (\bar{\psi}\psi)^2\right)\end{align}
and introduces the Hubbard-Stratanovich field $\sigma$ as
\begin{align}
S=\int d^dx\,\left(\bar{\psi}\slashed{\pl}\psi+\frac{\sigma}{\sqrt{N}}(\bar{\psi}\psi)- \frac12\sigma^2\right)\,.
\label{SGN-HS}
\end{align}
The auxiliary field acquires an induced (non-local) kinetic term via fermion loops
\begin{equation}
S_{\sigma} = -\frac{1}{2}\int d^d x d^d y \sigma(x)\sigma(y)\langle \frac{1}{\sqrt{N}}\bar\psi\psi(x) \frac{1}{\sqrt{N}}\bar\psi\psi(y)\rangle_0 +O(1/N)\,,
\end{equation}
where we have dropped the quadratic term in (\ref{SGN-HS}) as it does not contribute in the UV limit.
Inverting the induced quadratic term, one finds the $\sigma$ 2-point function to leading order in $1/N$ to be
\begin{align}
\langle \sigma(x_1) \sigma(x_2) \rangle&=\frac{C_{\sigma\sigma}}{x_{12}^2}\,,\qquad C_{\sigma\sigma}=-\frac{2 (d-2)  \Gamma (d-1)\sin \left(\frac{\pi  d}{2}\right)}{\pi  \Gamma \left(\frac{d}{2}\right)^2}\,,
\label{sig-2pt}
\end{align}
so that $\sigma \sim \bar\psi\psi$ is a scalar primary with $\Delta=1+O(1/N)$ at the UV fixed point.

The anomalous dimensions of $\psi$ and $\sigma$ to leading order in the $1/N$ expansion are well-known
\cite{Gracey:1992cp,Gracey:1990wi,Vasiliev:1992wr,Gracey:1993kc}:\footnote{$\Delta_\psi$ is known up to $1/N^3$ \cite{Vasiliev:1992wr,Gracey:1993kc} and $\Delta_\sigma$ up to $1/N^2$ \cite{Gracey:1992cp,Vasiliev:1992wr}.}
\begin{align}
\label{leading1N}
\Delta_\psi&=\frac{d-1}2-\frac1{N}\frac{(d-2)\Gamma(d-1)\sin \left(\frac{\pi  d}{2}\right)}{\pi  d\Gamma \left(\frac{d}{2}\right)^2}\,,\\
\Delta_\sigma&=1+\frac1{N}\frac{4\Gamma(d)\sin \left(\frac{\pi  d}{2}\right)}{\pi d \Gamma \left(\frac{d}{2}\right)^2}.
\end{align}
Since $\psi$ is a nearly free field at large $N$, the leading anomalous dimension of $\psi$
can be readily obtained using the equations of motion
\begin{eqnarray}
\label{largeNeq}
\slashed{\pl} \psi = - \frac{1}{\sqrt{N}} \psi \sigma\,,\qquad\qquad
\pl^\mu \bar{\psi} \gamma_\mu = + \frac{1}{\sqrt{N}} \bar{\psi} \sigma,
\end{eqnarray}
in the spirit of \cite{Rychkov:2015naa}, but with $1/N$ playing the role of the small parameter
(a similar calculation in the large $N$ scalar CFT was carried out in \cite{Skvortsov:2015pea, Giombi:2016hkj}). In the interacting theory,
the fermion $\psi$ must have the two-point function
\begin{align}
\langle\psi^i \bar{\psi}_j\rangle=\delta^{i}_j\frac{C_{\psi\psi}\slashed{x_{12}}}{(x^2_{12})^{\Delta_\psi+\tfrac12}}
\end{align}
with $\Delta_{\psi} = (d-1)/2+\gamma_{\psi}$.
Applying the Dirac operators $\slashed{\pl}_1$ and $\slashed{\pl}_2$ on this two-point function, one gets
\begin{align}
\slashed{\pl}_1\slashed{\pl}_2\langle\psi^i \bar{\psi}_j\rangle=-2\gamma_\psi(d+2\gamma_\psi)\delta^{i}_j\frac{C_{\psi\psi}\slashed{x_{12}}}{(x^2_{12})^{\Delta_\psi+\tfrac32}} \,,
\end{align}
which can be compared with the insertion of \eqref{largeNeq}
\begin{align}
-\frac{1}{N}\langle\psi^i\sigma\bar{\psi}_j\sigma\rangle=-\frac{1}{N}C_{\sigma\sigma} \delta^i_j \frac{C_{\psi\psi}\slashed{x_{12}}}{(x^2_{12})^{\Delta_\psi+\tfrac32}} \,.
\end{align}
This yields to the leading order in $1/N$:
\begin{align}
\label{psidim}
\gamma_\psi&=\frac{C_{\sigma\sigma}}{2d N}\,,
\end{align}
which, using (\ref{sig-2pt}), can be seen to be in full agreement with \eqref{leading1N}. Interestingly, the anomalous dimension of $\sigma$ can also be
reconstructed by using the equation of motion method, by considering the 3-point function $\langle \psi \bar\psi \sigma\rangle$. We will carry out
this calculation in section \ref{scalar-comp} below, and proceed here with the analysis of the weakly broken higher-spin operators.

To find the non-conservation operator for the higher-spin currents, we need to plug the equations of motion \eqref{largeNeq}
into the master formula for the non-conservation operator \eqref{nonconsgen}. As a result we find two type of terms:
with two gamma-matrices and without gamma-matrices:
\begin{align}
\label{Desc}
K_{s-1}= \frac{1}{\sqrt{N}}\Big(k_1 (\pl_i) \bar{\psi}(x_1)\psi(x_2)\sigma(x_3)+k_2 (\pl_i) \bar{\psi}(x_1)\gamma_{\mu\nu}\psi(x_2)\pl^\mu_3 z^\nu\sigma(x_3)\Big)
\end{align}
where
\begin{align}
    \begin{aligned}
    k_1 &\equiv \big[q_s^{d}(\deh_1+\deh_3,\deh_2) + (-1)^s q_s^d(\deh_2+\deh_3,\deh_1)\big] +\\
    &\qquad\qquad+(d-1)\deh_3 \big[q_{s-1}^{d+2}(\deh_1+\deh_3,\deh_2) + (-1)^s q_{s-1}^{d+2}(\deh_2+\deh_3,\deh_1)\big]\,,\\
    k_2&\equiv (d-1)\big[q_{s-1}^{d+2}(\deh_1+\deh_3,\deh_2) - (-1)^s q_{s-1}^{d+2}(\deh_2+\deh_3,\deh_1)\big]\,.
    \end{aligned}
\end{align}
The non-conservation operator $K_{s-1}$ must be a conformal primary of the unbroken theory\footnote{This can be seen by acting with the special
conformal generators on the non-conservation equation, see for instance \cite{Aharony:2011jz, MZ}.}
In particular it should be possible to decompose it as
\begin{align}
\label{Ndecomp}
K_{s-1}= \sum_{a,c} \mathcal{B}_{a,c}^{s}\deh^{a} (B^{s_1}_\mu \pl_3^\mu) \deh^c\sigma + \sum_{a,c} \mathcal{C}_{a,c}^{s} \deh^{a} (\pl^\mu B^{s_1}_\mu) \deh^{c}\sigma+\sum_{a,c}\mathcal{A}_{a,c}^{s}\deh^a(\bar\psi\psi)\deh^c\sigma \delta_{a+c,s-1}\,,
\end{align}
where the summation range and the spin $s_1$ of the operators $B^{s_1}_\mu$ in the sums are fixed by the spin and conformal dimension counting: $s_1+a+c+1=s$ in the first two sums and $a+c+1=s$ in the last one.

First of all, we observe that we need the hook currents, i.e. the ones with $(s,1)$ symmetry, as these contain two gamma matrices while the usual totally-symmetric currents do not contribute at all. It is also important to take the descendant $\pl^\mu B_\mu$ into account since it does not vanish.
The last terms involving the $\bar{\psi}\psi$-singlet appears only in the case of singlet currents of even spins. Note that at the large $N$ UV fixed point, this
term should be projected out due to the $\sigma$ equation of motion, which is formally $\bar\psi\psi=0$ ($\sigma$ replaces $\bar\psi\psi$ at the
UV fixed point).

Taking notice that the $\gamma$-part of the non-conservation operator is exactly like in the large-$N$ bosonic model \cite{Skvortsov:2015pea, Giombi:2016hkj},
we immediately find
\begin{align}
\mathcal{B}_{a,c}^{s}&=-\frac{2 (a+c+1)! (a+c-\nu -s+1) (a+2 c+1-2 (s+\nu ))!}{a! c! (c+1)! (a+c-2 (s+\nu ))!}\,,\label{ClebshGordonN}
\end{align}
which is assumed to vanish for $a+c$ even. The formula works both for even and odd spins.
Note that everything depends on $s+\nu$ only ($\nu=(d-3)/2$). Analogously,
\begin{align*}
\mathcal{C}_{a,c}^{s}&=\frac{2 (a+c+1)! (-a-c+\nu +s-1) (a+c-2 (\nu +s)+1) (a+2 c-2 (\nu +s))!}{a! (c!)^2 (a+c-s+1) (a+c-2 (\nu +s))!}\,,
\end{align*}
which is assumed to vanish for $a+c$ even or $a+c>s-2$. The only new part is due to the $(\bar\psi\psi)$-terms
\begin{align}
\mathcal{A}_{a,c}^{s}&=\frac{\left(\nu +\frac{s}{2}\right)^2 (2 \nu +s-1)!}{(a+2 \nu +1)!}\frac{4 (-1)^a s \left((1-s)_a\right){}^2}{\Gamma (a+1) \Gamma (s)}\,,
\end{align}
which is assumed to vanish for $s$ odd or unless $a+c=s-1$. The fact that the decomposition (\ref{Ndecomp})
is possible is a check of the non-conservation operator (\ref{Desc}).

\paragraph{Examples.} Let us consider a few explicit low spin examples.
The first nontrivial example is the spin-two singlet current, for which we find (omitting an overall factor)
\begin{align}
K^{s=2}&\sim(\deh_1+\deh_2+(1-d)\deh_3)\bar{\psi}\psi\sigma=(\pl(\bar{\psi}\psi)\sigma+(1-d)(\bar{\psi}\psi)\pl\sigma)\,,
\end{align}
which is conserved upon projecting out $\psib \psi$, the operator that is replaced by $\sigma$ in the large-$N$ treatment.
The spin-three non-conservation contains an anti-symmetric tensor:
\begin{align*}
K^{s=3}&=2(\deh_1-\deh_2)(-\deh_1-\deh_2+(d+1)\pl_3)\bar{\psi}\psi\sigma+(2(\deh_1+\deh_2))-d\pl_3)\bar{\psi}\gamma_{ab}\psi z^a\pl^b\sigma\\
&=\left[d B_{a,u}\pl^u\pl_a\sigma-2\pl_a B_{a,u}\pl^u\sigma-2(d+1)\pl^uB_{a,u}\pl_a\sigma+2\pl_a\pl^uB_{a,u}\sigma\right]z^az^a\,.
\end{align*}
The spin-four non-conservation contains a genuine mixed-symmetry current:
\begin{align*}
K^{s=4}
&=\left[5 B_{aa,u}\pl^u \pl_a\sigma -2\pl_a B_{aa,u}\pl^u\sigma-6\pl^uB_{aa,u}\pl_a\sigma+\pl_a\pl^uB_{aa,u}\sigma\right]z^az^az^a\\
&+\left[\frac83(\bar{\psi}\psi)\pl_a\pl_a\pl_a\sigma-12\pl_a(\bar{\psi}\psi)\pl_a\pl_a\sigma+8\pl_a\pl_a(\bar{\psi}\psi)\pl_a\sigma-\frac23\pl_a\pl_a\pl_a(\bar{\psi}\psi)\sigma\right]z^az^az^a
\end{align*}
where the formula is written in $d=3$ to simplify the coefficients and the last line displays the contribution of the $\bar{\psi}\psi$ operator that needs to be dropped for the singlet currents. As a result one sees the formula from \cite{Maldacena:2012sf, Giombi:2016zwa}.

In principle, one can use the decomposition \eqref{Ndecomp} to directly compute the two-point function in \eqref{mainform} and thus the anomalous dimensions.
However, the sums which appear are quite involved to calculate, and in practice we find more convenient to compute $\langle K_{s-1}K_{s-1}\rangle$
directly in terms of the form (\ref{Desc}).
To do this, one can start with the Schwinger representation (\ref{schw}).
Note that, using this representation, we can trade derivatives at point $0$ in \eqref{mainform} to $x$-derivatives by flipping their signs. The action of the projected derivatives on the integral is trivial: $\deh^n e^{-\alpha x^2} = (-2\alpha\hat{x})^n e^{-\alpha x^2}$ since $\deh \hat{x}=0$. Owing to this, the differential operators in the descendant
are replaced with a polynomial in $\alpha$ parameters. The derivatives which are not contracted with the null polarization vector require a bit more work,
but after some manipulations it is not difficult to see that the calculation eventually reduces to evaluating some integrals over $\alpha$ parameters,
which can be performed using the properties of Gegenbauer polynomials. Following this procedure to compute $\langle K_{s-1} K_{s-1}\rangle$, and using
the master formula (\ref{mainform}), we arrive at the following result
\begin{align}
\label{1Nnonsinglet}
    \gamma_s^A =   2\gamma_{\psi} \left( 1-\frac{(d-2) d}{4(s+\frac{d}{2}-2) (s+\frac{d}{2}-1)}\right) \,,
    \end{align}
which is valid for the non-singlet currents of all spins (and for odd spin singlets, which coincide with odd spin non-singlets).
To the best of our knowledge, this result was first obtained
in \cite{Muta:1976js} (using a standard Feynman diagram approach).

For singlet currents of even spins, the above result is not correct, because one has to subtract by hand the piece of the descendant (\ref{Ndecomp})
involving the scalar singlet $\bar\psi\psi$, as explained above. Subtracting from (\ref{1Nnonsinglet}) the contribution of the last term in (\ref{Ndecomp})
\begin{align}
\label{j0-piece}
\sum_{a,b} \mathcal{A}_{a,s-1-n}\mathcal{A}_{b,s-1-b} (\deh)^{a+b}(-1)^b \langle \bar{\psi}\psi \bar{\psi}\psi\rangle (\deh)^{2(s-1)-a-b}(-1)^{s-1-b} \langle\sigma \sigma\rangle
\end{align}
we obtain the final result for even spin singlets:
\begin{align}
\label{1Nsinglet}
    \gamma_s=2\gamma_{\psi}  \left( 1-\frac{(d-2) d}{4(s+\frac{d}{2}-2) (s+\frac{d}{2}-1)} - \frac{\Gamma(d+1)\Gamma(s+1)}{2(d-1)(s+\frac{d}{2}-2) (s+\frac{d}{2}-1)\Gamma(d+s-3)} \right)\,,
\end{align}
where the last term is due to the subtraction of (\ref{j0-piece}). This result is again in agreement with \cite{Muta:1976js}. Interestingly, the structure
of this result, as well as (\ref{1Nnonsinglet}), is identical to the ones in the critical large $N$ scalar model
\cite{Lang:1992zw, Skvortsov:2015pea, Giombi:2016hkj,Hikida:2016cla}, up to overall factor of $\gamma_{\psi}$ replacing $\gamma_{\phi}$. In particular, note that
the large spin expansion of (\ref{1Nsinglet}) reads
\begin{equation}
\gamma_s = 2\gamma_{\psi} \left(1-\frac{(d-2) d}{4}\frac{1}{s^2}-\frac{\Gamma\left(d+1\right)}{2(d-1)}\frac{1}{s^{d-2}}+\ldots \right)\,.
\end{equation}
The leading spin independent term agrees, of course, with the general expectation that $\Delta_s \sim 2\Delta_{\psi}+s$ at large spin.
Naively, following the arguments in \cite{Alday:2007mf, Fitzpatrick:2012yx, Komargodski:2012ek}, one might have expected a term of order $1/s$ corresponding
to the exchange of the scalar operator $\sigma$ with $\Delta=1+O(1/N)$, however we see that such term is absent.
Expanding at large spin the recent $1/N^2$ result in \cite{Manashov:2016uam}, we find that the anomalous dimensions include terms of order
 $1/s^{d-2}$, $1/s^2$, $\log(s)/s^{d-2}$ and $\log(s)/s^2$, but no terms of order $1/s$.
It would be interesting to understand this by generalizing the analysis of
\cite{Fitzpatrick:2012yx, Komargodski:2012ek, Alday:2015ewa}, or the approach of \cite{Alday:2016njk, Alday:2016jfr},
to the case of 4-point functions of fermionic operators.

\subsection{Gross-Neveu in $d=2+\epsilon$}
\label{sec:Gross-Neveu in 2+}
The Gross-Neveu model in dimension $d$ is defined by the action:
\begin{align}
S&=\int d^dx\, \left(\bar{\psi}\slashed{\pl}\psi +\frac12 g(\bar{\psi}\psi)^2\right)
\end{align}
where $\bar{\psi} \psi\equiv \bar{\psi}^i \psi_i$ and the action enjoys $U(N_f)$-symmetry.
The one-loop results for the $\beta$-function and the anomalous dimensions of the lowest lying operators $\psi$ and $\bar{\psi}\psi$
are well known (see e.g. \cite{Moshe:2003xn} for a review):
\begin{align}
\beta&=\epsilon g-(N-2) \frac{g^2}{2\pi}+...\,, && g_*=\frac{2\pi}{N-2}\epsilon\,,\\
\gamma_\psi&=\frac{N-1}{16\pi^2}g^2=\frac{N-1}{4(N-2)^2}\epsilon^2\,, && \Delta_\psi=\frac{d-1}2+\gamma_\psi\,,\\
\gamma_{\psi^2}&=-\frac{1}{2\pi}g=-\frac{\epsilon}{N-2}\,, && \Delta_\phi=d-1+\gamma_{\psi^2}\,.
\end{align}
Note that $N$ here is $ N_f \text{tr} \boldsymbol{1}$ the total number of the field components. The equations of motion take the following form
\begin{eqnarray}
\slashed{\pl} \psi = - g \psi (\bar{\psi} \psi)\,,\qquad\qquad \pl^\mu \bar{\psi} \gamma_\mu = + g \bar{\psi} (\bar{\psi} \psi)\,.
\end{eqnarray}
Following similar methods as in \cite{Rychkov:2015naa}, it is possible to use these equations to derive the above anomalous dimensions,
as well as the dimension of higher order composite scalar operators \cite{Raju:2015fza, Ghosh:2015opa}. Here, we use a similar approach to fix the leading
order anomalous dimensions of the weakly broken higher-spin currents.

The calculation of the divergence of the currents follows similar steps as in the previous sections. We find
\begin{align}
\label{DescGN}
K_{s-1}= g\Big(k_1 (\pl_i) \bar{\psi}(x_1)\psi(x_2)\bar{\psi}(x_3)\psi(x_4)+k_2 (\pl_i) \bar{\psi}(x_1)\gamma_{\mu\nu}\psi(x_2)(\pl^\mu_3+\pl^\mu_4 ) z^\nu\bar{\psi}(x_3)\psi(x_4)\Big)
\end{align}
where
\begin{equation}
    \begin{aligned}
    k_1 &\equiv \big[q_s^{2}(\deh_1+\deh_3+\deh_4,\deh_2) + (-1)^s q_s^2(\deh_2+\deh_3+\deh_4,\deh_1)\big] \\
  &+ \deh_3 \big[q_{s-1}^{4}(\deh_1+\deh_3+\deh_4,\deh_2) + (-1)^s q_{s-1}^{4}(\deh_2+\deh_3+\deh_4,\deh_1)\big]\,,\\
    k_2&\equiv  \big[q_{s-1}^{4}(\deh_1+\deh_3+\deh_4,\deh_2) - (-1)^s q_{s-1}^{4}(\deh_2+\deh_3+\deh_4,\deh_1)\big]\,.
    \end{aligned}
\end{equation}
To find the anomalous dimension according to the formula \eqref{mainform},
we have to calculate a two-point function of four fermionic operators at points $x$ and $0$.
There are three different ways to contract the spinor and $U(N_f)$ indices, pair by pair or ``threading" through all 8 operators.
The first diagram, obtained by contracting the first pair at point $x$ with the corresponding one at point $0$, gives a contribution
\begin{equation}
  \frac{N \epsilon^2}{2 (N-2)^2} \,.
\end{equation}
The second diagram is obtained by contracting the first pair at point $x$ with the second pair at point $0$. It is non-zero only for even singlets
and gives a contribution to $\gamma_s$
\begin{equation}
   -\frac{1+(-1)^s}{2}\frac{N \epsilon^2}{2 (N-2)^2} \,.
\end{equation}
Finally, the third diagram is the one where spinor and $U(N_f)$ indices make a single loop threading all $8$ fermions, and yields a contribution
\begin{equation}
 -\frac{1-(-1)^s}{2}\frac{ \epsilon^2}{(N-2)^2} \,.
\end{equation}
Summing these results up we get for the non-singlets:
\begin{equation}
\begin{aligned}
  \gamma_{s}^A &= \frac{N \epsilon^2}{2 (N-2)^2}\,,\qquad  s{~~\rm even}\,,\\
  \gamma_{s}^A &= \frac{(N-2) \epsilon^2}{2 (N-2)^2}=\frac{ \epsilon^2}{2 (N-2)}\,,\qquad  s>1{~~\rm odd}\,,
\end{aligned}
\end{equation}
and $\gamma^A_{s=1}=0$ in accordance with the exact $U(N_f)$ symmetry. For the even spin singlets, we obtain\footnote{The result for odd spin singlets is
the same as for odd spin non-singlets.}
\begin{align}
  \gamma_{s} &= O(\epsilon^3)\,,\qquad  s \ge 2{~~\rm even} \,.
\end{align}

Note that, to this order, all these results are independent of the spin (except for the vanishing of $\gamma^A_{s=1}$).
One can check that they match precisely the expansion of the $1/N$ values \eqref{1Nnonsinglet} and \eqref{1Nsinglet} near $d=2+\epsilon$.
Including also the recent $1/N^2$ results in \cite{Manashov:2016uam}, we find that the $2+\epsilon$ expansion of the available large $N$ results
is
\begin{eqnarray*}
\gamma^A_{s>1} &=&\epsilon^2\left(\frac{1}{2N} + \frac{3+(-1)^s}{2N^2}\right)
+\epsilon^3\left(
\frac{\frac{1}{s-s^2}-1}{4N}
+\frac{1+\frac{2}{s-s^2}-2 (-1)^s (\psi(s)+\gamma)}{4 N^2}\right)+O(\epsilon^4), \cr
\gamma_s &=& \epsilon^3 \left(\frac{4}{s}-\frac{4}{s-1}-2+4 (\psi(s)+\gamma )\right)\left(\frac{1}{8N}+\frac{3}{8N^2}+\ldots\right)
+O(\epsilon^4)\,,\qquad s{~~\rm even}\,,
\end{eqnarray*}
where $\psi(s)$ is the digamma function and $\gamma$ the Euler-Mascheroni constant. Note that to leading order in $\epsilon$ these precisely agree
with the results derived above. Note also that a non-trivial spin dependence, including terms of order $\sim \log(s)$ at large $s$,
appears at the next-to-leading order in the $\epsilon$-expansion.

\subsection{Gross-Neveu-Yukawa in $d=4-\epsilon$}
\label{sec:GNY}
The Gross-Neveu-Yukawa (GNY) action is defined as:
\begin{align}
S=\int d^dx\,\left(\bar{\psi}_i(\slashed{\pl}+g_1\sigma )\psi^i+ \frac12(\partial_{\mu} \sigma)^2+ \frac{g_2}{24} \sigma^4\right)
\end{align}
This model has a perturbative IR fixed point in $d=4-\epsilon$ which is in the same universality class as the UV fixed point of the GN model,
and thus provides a different description of the same interacting fermionic CFT \cite{Hasenfratz:1991it, ZinnJustin:1991yn, Moshe:2003xn}. The one-loop beta functions
yield the fixed point couplings
\begin{equation}
\begin{aligned}
    &(g^{*}_1)^2 =\frac{16 \pi^2 \epsilon}{N+6}\,,\qquad g^{*}_2 = 16 \pi^2 R \epsilon\,, \\
    &R =\frac{24N}{(N+6)[ (N-6)+\sqrt{N^2+132N+36}]}\,,
\end{aligned}
\end{equation}
and the leading anomalous dimensions of $\psi$ and $\sigma$ are
\begin{eqnarray}
\label{GNYdata}
\gamma_{\psi} = \frac{(g^{*}_1)^2}{32\pi^2}=\frac{\epsilon}{2(N+6)}\,,\\
\gamma_{\sigma} = \frac{N(g^{*}_1)^2}{32\pi^2} = \frac{N\epsilon}{2(N+6)}\,.
\end{eqnarray}
The equations of motion are
\begin{equation}
\begin{aligned}
\label{eom-GNY}
&\slashed{\partial} \psi^i = -g_1 \sigma \psi^i\,, \\
&\partial^2 \sigma =  g_1 \bar{\psi}_i \psi^i + \frac{g_2}{6} \sigma^3\,.
\end{aligned}
\end{equation}
From the abstract CFT point of view, these equations describe the multiplet recombination which makes $\psi$ and $\sigma$ into long representations of the conformal algebra
with $\Delta_{\psi}>(d-1)/2$ and $\Delta_{\sigma} > d/2-1$, and can be used to fix the leading order anomalous dimensions. Since the $\psi$ equation of motion formally coincides with the large-$N$ equation \eqref{largeNeq}, one arrives at the same relation (\ref{psidim}), with $d=4$:
\begin{align}
\label{gpsi-eom}
\gamma_\psi&=\frac{1}{8}C_{\sigma\sigma}(g^{*}_1)^2 = \frac{(g_1^*)^2}{32\pi^2}\,,
\end{align}
where we have used $C_{\sigma\sigma} = \frac{1}{4 \pi^2}$, since $\sigma$ is now a canonically normalized (near) free field.
As for the $\sigma$ anomalous dimension, one can see that to leading order
in the breaking parameter we can neglect the $\sigma^3$ term in (\ref{eom-GNY}), and by a calculation similar to the one in the scalar $\phi^4$ theory \cite{Rychkov:2015naa, Skvortsov:2015pea, Giombi:2016hkj}, one finds (setting $d=4$ which is appropriate to leading order):
\begin{equation}
\gamma_{\sigma} =\frac{1}{32C_{\sigma\sigma}} (g_1^*)^2 x_{12}^6 \langle \bar\psi \psi(x_1) \bar\psi\psi(x_2) \rangle_0   \,,
\end{equation}
which yields
\begin{equation}
\label{gsig-eom}
\gamma_{\sigma} = \frac{N C_{\psi\psi}^2(g_1^*)^2}{32C_{\sigma\sigma}}=\frac{N(g_1^*)^2}{32\pi^2}\,,
\end{equation}
in agreement with (\ref{GNYdata}). The relation between $g_1^*$ and $\epsilon$ can also be reconstructed without input from Feynman diagrams and beta functions by
applying the equations of motion (\ref{eom-GNY}) to the 3-point function $\langle \psi \bar\psi \sigma\rangle$, as will be explained in section \ref{scalar-comp} below.

Let us now turn to the weakly broken higher-spin currents of the model (focusing on the totally symmetric operators only). By applying the general non conservation formula \eqref{nonconsgen}, the divergence of the fermion bilinear currents (flavor indices omitted)
\begin{equation}
\hat{J}_{s,\psi} =  (\deh_1+\deh_2)^{s-1}C^{3/2}_{s}\Big(\frac{\deh_1-\deh_2}{\deh_1+\deh_2}\Big)\bar\psi(x_1) \hat\gamma \psi(x_2)\Big{|}_{x_1,x_2\rightarrow x}
\end{equation}
is found to be
\begin{align}
\label{DescGNY}
\hat{K}_{s-1, \psi}= g_1\Big(k_1 (\pl_i) \bar{\psi}(x_1)\psi(x_2)\sigma(x_3)+k_2 (\pl_i) \bar{\psi}(x_1)\gamma_{\mu\nu}\psi(x_2)\pl^\mu_3 z^\nu\sigma(x_3)\Big)
\end{align}
and is almost identical to the large-$N$ limit, with $k_1$ and $k_2$ given now for $d=4$ by
\begin{align}
    \begin{aligned}
    k_1 &\equiv \big[q_s^{4}(\deh_1+\deh_3,\deh_2) + (-1)^s q_s^4(\deh_2+\deh_3,\deh_1)\big] + \\
    &\qquad\qquad+3\deh_3 \big[q_{s-1}^{6}(\deh_1+\deh_3,\deh_2) + (-1)^s q_{s-1}^{6}(\deh_2+\deh_3,\deh_1)\big]\,,\\
    k_2&\equiv 3 \big[q_{s-1}^{6}(\deh_1+\deh_3,\deh_2) - (-1)^s q_{s-1}^{6}(\deh_2+\deh_3,\deh_1)\big]\,.
    \end{aligned}
\end{align}
For the non-singlets and odd spin singlets, the fermionic bilinear currents do not mix with other operators, so the calculation of the anomalous dimensions proceeds exactly
as in the $1/N$ expansion, with $C_{\sigma\sigma}$ in the latter replaced now by $\frac{1}{4\pi^2}(g^{*}_1)^2$, and we find
\begin{align}
    \gamma_s^A =\frac{(g^*_1)^2}{16 \pi^2}\frac{(s-1)(s+2)}{s(s+1)}= 2\gamma_{\psi} \left(1-\frac{2}{s(s+1)}\right)\,.
\end{align}
Note that this vanishes for $s=1$, as it should. Expanding this result at large $N$ to the order $1/N^2$, we find agreement with both \eqref{1Nnonsinglet} and the result of \cite{Manashov:2016uam}
\be
\gamma_s^A = \epsilon \left(\frac{1}{N}-\frac{6}{N^2}+\ldots \right)\left(1-\frac{2}{s(s+1)}\right)+O(\epsilon^2)\,.
\ee

In the case of even spin singlets, a novelty compared to the large $N$ calculation is the appearance of mixing with the scalar bilinear higher-spin currents
\begin{align}
    \hat{J}_{s,\sigma} = (\deh_1+\deh_2)^{s}C^{1/2}_{s}\Big(\frac{\deh_1-\deh_2}{\deh_1+\deh_2}\Big)\sigma (x_1) \sigma (x_2)\Big{|}_{x_1,x_2\rightarrow x} \,,\qquad s=2,4,6,\ldots
\end{align}
The divergence of these currents, which we denote $\hat{K}_{s-1,\sigma}$, has two pieces according to equations of motion. However, to lowest order in $\epsilon$ we may ignore the piece coming with $g_2$ since $g^{*}_{1} \sim \sqrt{\epsilon}$ whereas $g^{*}_{2} \sim \epsilon$. To the lowest order, the descendant is then:
\begin{align}
    \hat{K}_{s-1,\sigma} = 2 g_1  q^{4}_{s} (\deh_1+\deh_2,\deh_3) \bar{\psi}_i(x_1) \psi^i(x_2) \sigma(x_3)\,.
\end{align}
It is evident that this term induces the mixing between the $\hat{J}_{s,\psi}$ and $\hat{J}_{s,\sigma}$ since there are non-zero off-diagonal 2-point functions
$\langle \hat K_{s-1,\sigma} \hat K_{s-1,\psi}\rangle$.\footnote{A similar mixing of two towers of nearly conserved higher-spin currents takes place
in the cubic $O(N)$ scalar field theory in $d=6-\epsilon$ \cite{Giombi:2016hkj}.}
The calculation can be carried out using the Schwinger representation of the propagator, and using (\ref{mainform}) we find the mixing matrix
\begin{equation}
\frac{(g^*_1)^2}{16 \pi^2}\begin{bmatrix}
   \frac{(s-1)(s+2)}{s(s+1)} & \frac{-2 \sqrt{N}}{\sqrt{s(s+1})}   \\
  \frac{-2 \sqrt{N}}{\sqrt{s(s+1})} &  N
\end{bmatrix}\,.
\label{mixingM}
\end{equation}
This leads to the anomalous dimensions:
\begin{align}
    \gamma_s^{\pm} = \frac{g^2_1}{16\pi^2} \frac{-2 + (N+1) s (1 + s) \pm \sqrt{
  4 + s (1 + s) (-4 + 20 N + ( N-1)^2 s + ( N-1)^2 s^2)}}{
 2 s (1 + s)}
\label{gammPm}
\end{align}
The eigenvalue $\gamma_s^{-}$ corresponds to the tower of weakly broken currents with twist near $d-2$ which is present in the large $N$ treatment of the CFT. Indeed, one may check that $\gamma_{s=2}^{-}=0$, indicating that the corresponding eigenstate is the conserved stress tensor of the CFT. The large $N$ expansion of the
anomalous dimensions reads
\begin{align}
    \gamma_s^{-} = \frac{\epsilon}{N} \frac{(s-2)(s+3)}{s(s+1)} -\frac{2\epsilon}{N^2} \frac{(s-2)(s+3)(3s^2+3s+2)}{s^2(s+1)^2} +\ldots \,,
\end{align}
which again agrees with the expansion around $d=4-\epsilon$ of the $1/N$ value \eqref{1Nsinglet} and $1/N^2$ results from \cite{Manashov:2016uam}.

The second eigenvalue $\gamma_s^{+}$ should instead correspond in the large $N$ approach to the ``double-trace" operators $ \sigma \partial^s \sigma \sim (\bar{\psi}\psi) \partial^s (\bar{\psi}\psi)$. This is suggested by expanding (\ref{gammPm}) at large $N$, which yields
\begin{align}
\label{sigma4}
    \Delta^{+}= d - 2 + s+ \gamma_s^{+} = 2+s  -2\frac{\epsilon}{N}\frac{ 3 s^2+3s -2}{s (1 + s)}+O(1/N^2)\,,
\end{align}
corresponding to an operator with twist $2+O(1/N)$. In section \ref{sigmadsigma}, we will compute the scaling dimensions of the $\sigma \partial^s\sigma$ operators in the
large $N$ theory as a function of $d$, and explicitly verify agreement with (\ref{sigma4}). This provides a non-trivial test of the identification of the IR fixed point of the GNY model with the UV fixed point of the GN model.

Let us also give the large spin expansion of the anomalous dimensions (\ref{gammPm}). Writing the result in terms of $\gamma_{\psi}$ and $\gamma_{\sigma}$, we find
\begin{equation}
\begin{aligned}
&\gamma_s^-= 2\gamma_{\psi}\left(1-\frac{6 N-2}{(N-1) s^2}+\frac{6 N-2}{(N-1) s^3}+\ldots\right)\,,\\
&\gamma_s^+= 2\gamma_{\sigma}\left(1+\frac{4}{(N-1) s^2}-\frac{4}{(N-1) s^3}+\ldots\right)\,.
\end{aligned}
\end{equation}
Note the absence of $1/s$ terms, as observed above in the large $N$ results.

The case $N=1$ is special and deserves a separate comment. Since $N=N_f\, \rm tr 1$, this corresponds to a formal analytic continuation which should yield a CFT with
a single 2-component Majorana fermion in $d=3$. It has been argued that the IR fixed point in this case displays emergent supersymmetry \cite{Thomas, Grover:2013rc, Bashkirov:2013vya, Shimada:2015gda, Iliesiu:2015qra, Fei:2016sgs}. Indeed, we can see that
for $N=1$ the dimension of $\sigma$ and $\psi$ in (\ref{GNYdata}) coincide. Further checks, including higher orders in the $\epsilon$-expansion, can be found in \cite{Fei:2016sgs}. Note that for $N=1$ the square roots in (\ref{gammPm}) simplify, and we get the results
\begin{equation}
\gamma_s^{-} =2\gamma_{\psi}\left(1-\frac{2}{s}\right) \,,\qquad \gamma_s^{+} = 2\gamma_{\psi}\left(1+\frac{2}{s+1}\right) \,.
\end{equation}
It would be interesting to study the half-integer higher-spin operators $\sim \sigma \partial^{s-1/2} \psi$, and check how the operators organize into supersymmetric multiplets.

\section{Composite operators in fermionic and scalar CFTs}
In this section we will calculate the dimensions of various composite operators in fermionic $U(N)$ theories and also in the bosonic $O(N)$ model. Mostly, we will be interested in the large-$N$ results for operators built from the auxiliary field $\sigma$: $\sigma$ itself, $\sigma^k$ and $\sigma \pl^s \sigma$.

\subsection{Some scalar operators}
\label{scalar-comp}
Let us consider the traceless (adjoint) scalar operator $(O^A)_{j}^{i}=\bar{\psi}_{j} \psi^i - \frac{\delta_j^i}{N}  \bar{\psi} \psi$ in
the large $N$ critical fermion theory. Its leading order anomalous dimension can be fixed by starting with the 3-point function
\be
\label{psipsiO}
\langle \psi^i (x_1) \bar{\psi}_j (x_2)(O^A)_k^l (x_3)  \rangle = (\delta_j^l \delta_k^i - \frac{1}{N} \delta_j^i\delta_k^l) \frac{C_{\psi \psi O} \slashed{x_{13}} \slashed{x_{23}}}{x_{12}^{2\Delta_\psi-\Delta_O+2}x_{13}^{\Delta_O+1}x_{23}^{\Delta_O+1}}\,,
\ee
where the structure on the right-hand side is completely fixed by conformal symmetry, for general $\Delta_{\psi}$ and $\Delta_O$. Acting with the Dirac operators $\slashed{\pl}_1 \slashed{\pl}_2$ on the right-hand side, we find
\be
\label{direct}
(2\Delta_\psi +\Delta_O-d) (2\Delta_\psi-\Delta_O)(\delta_j^l \delta_k^i - \frac{1}{N} \delta_j^i\delta_k^l) C_{\psi \psi O}\frac{\slashed{x_{13}} \slashed{x_{23}}}
{x_{12}^{2\Delta_\psi-\Delta_O}x_{13}^{\Delta_O+1}x_{23}^{\Delta_O+1}}+\ldots \,,
\ee
where we have omitted terms proportional to the identity in the spinor space, which will not play a role in the following leading order calculations. After writing $\Delta_\psi = d/2 -1/2+\gamma_\psi$, $\Delta_O = d-1 + \gamma_O$, and expanding to leading order in the anomalous dimensions, the prefactor above becomes
\be
(d-2) (2\gamma_\psi-\gamma_O)
\ee
and hence this correlator can be used to fix $\gamma_O$ in terms of $\gamma_{\psi}$. Indeed, inserting the equation of motion $\slashed{\pl} \psi =-\frac{\sigma \psi}{\sqrt{N}}$
on the left-hand side of (\ref{psipsiO}), we get, to leading order at large $N$
\be
\begin{aligned}
&-\frac{1}{N} \langle \psi^i \sigma(x_1) \bar{\psi}_j \sigma(x_2) (O^A)_k^l(x_3) \rangle = -\frac{1}{N} \langle \sigma(x_1) \sigma(x_2)  \rangle \langle  \psi^i (x_1) \bar{\psi}_j(x_2)(O^A)_k^l (x_3) \rangle= \\
&= -\frac{1}{N} (\delta_j^l \delta_k^i - \frac{1}{N} \delta_j^i\delta_k^l) \frac{C_{\sigma \sigma} C_{\psi \psi O}  \slashed{x_{13}} \slashed{x_{23}}}{x_{12}^2 x_{13}^d x_{23}^d}\,.
\end{aligned}
\ee
This gives:
\be
\gamma_O= 2 \gamma_\psi +\frac{C_{\sigma\sigma}}{N(d-2)} = \frac{4(d-1)}{d-2}\gamma_\psi\,.
\ee
In $d=3$, this agrees with the result obtained in \cite{Iliesiu:2015qra} via a standard Feynman diagram calculation. As far as we know, the general $d$ result was not given elsewhere in the literature.

The same calculation can be applied in the case of the GNY model in $d=4-\epsilon$, where following the similar steps we arrive at
\be
\gamma_O = 2 \gamma_\psi + C_{\sigma\sigma} \frac{(g_1^*)^2}{d-2} =  \frac{3\epsilon}{N+6}\,.
\ee

Analogously, we can fix the anomalous dimension of the scalar singlet $\sigma$ starting from the 3-point function
\be
\langle \psi^i (x_1) \bar{\psi}_j (x_2) \sigma (x_3)  \rangle = \delta^i_j \frac{C_{\psi \psi \sigma} \slashed{x_{13}} \slashed{x_{23}}}{x_{12}^{2\Delta_\psi-\Delta_\sigma}x_{13}^{\Delta_\sigma+1}x_{23}^{\Delta_\sigma+1}}\,.
\ee
Direct application of two Dirac operators yields
\be
\label{direct-sig}
(d-2) (2\gamma_\psi + \gamma_\sigma) \delta^i_j \frac{C_{\psi \psi \sigma} \slashed{x_{13}} \slashed{x_{23}}}{x_{12}^{d}x_{13}^{2}x_{23}^{2}}+\ldots
\ee
where we used the formula \eqref{direct} with the $O$ replaced by $\sigma$, together with the expansion $\Delta_\sigma = 1+ \gamma_\sigma$.
Inserting equations of motion, one gets
\be
-\frac{1}{N} \langle  \psi^i \sigma(x_1) \bar{\psi}_j \sigma(x_2) \sigma(x_3) \rangle =- \frac{1}{N} \langle \sigma(x_1) \sigma(x_2) \rangle \langle  \psi^i (x_1) \bar{\psi}_j(x_2) \sigma (x_3) \rangle+\ldots  =-\frac{1}{N} \frac{C_{\sigma \sigma} C_{\psi\psi \sigma} }{x_{12}^{d}x_{13}^{2}x_{23}^{2}}+\ldots \,,
\ee
where we used that the three-point function  $\langle \sigma \sigma \sigma\rangle$ vanishes, and in the second step we have selected only the relevant contraction
that gives the tensor structure in (\ref{direct-sig}). The final result is then
\be
\gamma_\sigma = - 2 \gamma_\psi - \frac{C_{\sigma\sigma}}{N(d-2)}= - 4\gamma_\psi\frac{d-1}{d-2}\,,
\ee
which agrees with the result quoted earlier \eqref{leading1N}.
It is interesting to repeat the above calculation in the GNY model. Using \eqref{direct} and $\Delta_\sigma = d/2 -1 + \gamma_\sigma$, to leading order in the anomalous dimensions we find the relation
\be
2\left(2 \gamma_\psi + \gamma_\sigma-\frac{\epsilon}{2}\right) = -C_{\sigma \sigma} (g^*_1)^2 =-\frac{1}{4\pi^2}(g^*_1)^2\,.
\ee
Combining this with (\ref{gpsi-eom}) and (\ref{gsig-eom}) obtained from the $\psi$ and $\sigma$ 2-point functions, we see that we recover the correct fixed-point coupling given in (\ref{GNYdata}).

It should be possible by similar methods to fix the scaling dimensions of higher order composites of $\sigma$ and $\psi$. We will not pursue this in detail here, and just
use a shortcut to obtain the result for a simple class of operators, namely the composites $\sigma^k$ with twist $k+O(1/N)$ (see also \cite{Iliesiu:2015qra} for this
calculation in the case $d=3$). We note that the diagrammatic expansion suggests that, to leading order in $1/N$, $\Delta_{\sigma^k}$ is at most a quadratic function of $k$. More precisely, there are two diagrams, which connect the $\sigma$ legs pair by pair, so the $k$ dependence for these is $k(k-1)$. There are also leg corrections, which go like $k$. So, overall, we should have:
\be
\Delta_{\sigma^k} = k + \frac{Ak^2 + Bk}{N}+O(1/N^2)\,.
\ee
We can then use the known results for $k=1$ and $k=2$ \cite{Gracey:1990wi} to get:
\be
\label{delsigk-fermi}
\Delta_{\sigma^k} = k + \frac{2k(d-1)(d(k-1)-2)}{d-2}  \gamma_\psi\,.
\ee
Note that we may also write this result as
\be
\Delta_{\sigma^k}- k\Delta_{\sigma} =
\frac{k(k-1)}{2} \frac{4d(d-1)\gamma_{\psi}}{d-2}
= \frac{k(k-1)}{2}\left(\gamma_{\sigma^2}-2\gamma_{\sigma}\right)\,,
\ee
where the right-hand side has the interpretation of interaction energy for the multi-particle state in AdS which is dual to $\sigma^k$.

We may also apply the same method to the bosonic $O(N)$ model in the $1/N$ expansion, which can be developed
using the action for the scalar field $\phi^i$ and the auxiliary field $\sigma$:
\be
\label{actionlargeN}
S=\int d^d x\Big(  \frac{1}{2} \partial_{\mu} \phi_{i}  \partial^{\mu} \phi^{i}+ \frac{1}{2\sqrt{N}} \sigma \phi^i \phi^i\Big).
\ee
The $\sigma$ field acquires an effective non-local propagator upon integrating out $\phi^i$\footnote{The quadratic term in the resulting
$\sigma$ effective action is just proportional to the two-point function $\langle \phi^i\phi^i(x)\phi^j\phi^j(y)\rangle_0$ in the free theory.
The $\sigma$ propagator can be obtained by Fourier transforming to momentum space, inverting, and transforming back to $x$-space.}
\be
\langle \sigma(x_1) \sigma(x_2) \rangle = \frac{C_{\sigma\sigma}}{(x^2_{12})^2}\,,\qquad  C_{\sigma\sigma}=\frac{2^{d+2}\Gamma(\frac{d}{2}-\frac{1}{2})\sin(\pi d/2)}{\pi^{\frac{3}{2}}\Gamma(\frac{d}{2}-2)}\,,
\ee
so that $\sigma$, which replaces the scalar operator $\phi^i\phi^i$, is a primary operator of dimension $2+O(1/N)$ at the interacting fixed point.
Systematic perturbation theory can be developed using this effective propagator,
the canonical propagator  for $\phi^i$ and the $\sigma \phi^i\phi^i$ vertex, with $1/\sqrt{N}$ playing the role of the coupling constant. The equation of motion for $\phi$ is
\be
\label{largeNeqsc}
\partial^2 \phi^i = \frac{1}{\sqrt{N}}\sigma \phi^i
\ee
can be used \cite{Skvortsov:2015pea, Giombi:2016hkj} to readily derive the leading anomalous dimension of $\phi$
\be
\gamma_{\phi} = \frac{C_{\sigma\sigma}}{4Nd(d-2)}+O(1/N^2)\,.
\ee
Similarly to the fermionic calculation above, we can fix the anomalous dimension of the traceless scalar  $O^{ij}=\phi^{(i} \phi^{j)}-\frac{\delta^{ij}}{N} \phi^2$. One can study the following three-point function
\be
\langle \phi^i (x_1) \phi^j(x_2) O^{kl} (x_3) \rangle = (\delta^{ik}\delta^{jl} + \delta^{il} \delta^{jk} - \frac{2}{N} \delta^{ij} \delta^{kl})\frac{C_{\phi\phi O}}{x_{12}^{2\Delta_{\phi}-\Delta_O}x_{13}^{\Delta_O}x_{23}^{\Delta_O}}
\ee
as constrained by the conformal symmetry. The dimensions of operators are power series in $1/N$: $\Delta_{\phi}= d/2 - 1+\gamma_\phi$, $\Delta_O = d-2+ \gamma_O$. One can then act on this three-point function with the Laplacians $\square_1$ and $\square_2$ at points 1 and 2 respectively. Explicit differentiation gives, to the leading order in $1/N$:
\be
 2(d-2)(d-4)(2\gamma_{\phi} - \gamma_O) (\delta^{ik}\delta^{jl} + \delta^{il} \delta^{jk} - \frac{2}{N} \delta^{ij} \delta^{kl}) \frac{C_{\phi\phi O}}{x_{12}^{4} x_{13}^{d-2}x_{23}^{d-2}}\,,
\ee
where we replaced the dimensions of the operators with the tree-level values where possible. Now, inserting the equation of motion \eqref{largeNeqsc} in the correlation function, we get
\be
\begin{aligned}
&\frac{1}{N}\langle \sigma \phi^i (x_1) \sigma \phi^j (x_2) O^{kl} (x_3)\rangle = \frac{1}{N} \langle \sigma (x_1) \sigma (x_2) \rangle \langle \phi^i (x_1) \phi^j(x_2) O^{kl} (x_3) \rangle   \\
&= \frac{1}{N}(\delta^{ik}\delta^{jl} + \delta^{il} \delta^{jk} - \frac{2}{N} \delta^{ij} \delta^{kl})\frac{C_{\sigma\sigma} C_{\phi\phi O}}{x_{12}^{4} x_{13}^{d-2}x_{23}^{d-2}}
\end{aligned}
\ee
and so to the lowest order in $1/N$ we obtain:
\be
\gamma_O = 2\gamma_\phi - \frac{C_{\sigma\sigma}}{2N(d-2)(d-4)} = - \frac{8\gamma_\phi }{d-4} = -\frac{2^d\Gamma(\frac{d-1}{2})\sin(\frac{\pi d}{2})}{\pi^{\frac{3}{2}} \Gamma(\frac{d}{2}+1)}\,.
\ee

Now the calculation for the $\sigma$ operator is slightly more complicated although the general idea is the same. We begin with the three-point function:
\be
\langle \phi^i (x_1) \phi^j(x_2) \sigma (x_3) \rangle = \delta^{ij} \frac{C_{\phi\phi \sigma}}{x_{12}^{2\Delta_{\phi}-\Delta_\sigma}x_{13}^{\Delta_\sigma}x_{23}^{\Delta_\sigma}}\,.
\ee
Expanding $\Delta_\sigma = 2+ \gamma_\sigma$, and acting with the operators $\square_1$ and $\square_2$ gives:
\be
2\delta^{ij} (d-2)(d-4) (\gamma_\sigma+2\gamma_\phi) \frac{C_{\phi\phi\sigma}}{x_{12}^d x_{13}^2 x_{23}^2}+\ldots \,,
\label{sig-ON}
\ee
where we omitted two other terms which do not contain $\gamma_\sigma$, and have a different coordinate dependence. Inserting the equations of motion and working to leading order at large $N$, we get:
\be
\begin{aligned}
&\frac{1}{N} \langle \sigma \phi^i (x_1) \sigma \phi^j (x_2) \sigma (x_3)\rangle \\
&= \frac{1}{N} \left( \langle \sigma(x_1) \sigma(x_2) \rangle \langle \phi^i (x_1) \phi^j(x_2) \sigma (x_3) \rangle + \langle \phi^i(x_1) \phi^j(x_2) \rangle   \langle \sigma(x_1) \sigma(x_2) \sigma(x_3) \rangle+\ldots  \right)\\
&=\frac{1}{N}  \frac{\left( C_{\sigma \sigma} C_{\phi \phi \sigma} + C_{\phi \phi} C_{\sigma \sigma \sigma} \right) }{x_{12}^d x_{13}^2 x_{23}^2}+\ldots \,,
 \end{aligned}
\ee
where we have only kept the relevant structures that match (\ref{sig-ON}). Comparing the two results, we obtain the relation
\be
\gamma_\sigma = -2\gamma_\phi + \left(C_{\sigma \sigma} + \frac{C_{\phi \phi} C_{\sigma \sigma \sigma }}{C_{\phi\phi \sigma}}   \right) \frac{1}{2N(d-2)(d-4)}\,,
\ee
where $C_{\sigma \sigma \sigma}$ and $C_{\phi \phi \sigma}$ are the 3-point function coefficients, which both start at order $1/N$.
Their ratio is known to be, to leading order at large $N$ \cite{Petkou:1994ad}
\be
\frac{C_{\sigma\sigma\sigma}}{C_{\phi\phi\sigma}} = \frac{2(d-3) C_{\sigma \sigma}}{C_{\phi\phi}}\,,
\ee
so finally we find
\be
\gamma_\sigma= -2\gamma_\phi + \left(C_{\sigma \sigma} + 2(d-3)C_{\sigma\sigma}  \right) \frac{1}{2N(d-2)(d-4)} = \frac{4 (d-1)(d-2)}{d-4} \gamma_\phi\,,
\label{gamsig-boson}
\ee
which is the correct result \cite{Vasiliev:1982dc}.

The result for $\sigma^k$  operators for arbitrary $k$ was obtained in \cite{Lang:1992zw}, and is again constrained to be a quadratic function of $k$ to leading order in $1/N$. It reads
\be
\Delta_{\sigma^k}=2k+\frac{2k(d - 1) ((k - 1)d^2 + d + 4 - 3k d)}{d-4}\gamma_{\phi}+O(1/N^2) \,.
\ee
We can also express this result in terms of ``binding energies"
\be
\Delta_{\sigma^k}-k\Delta_{\sigma} =\frac{k(k-1)}{2} \frac{4d(d-1)(d-3)\gamma_{\phi}}{4-d}=\frac{k(k-1)}{2} \left(\gamma_{\sigma^2}-2\gamma_{\sigma}\right)\,.
\ee
Quite interestingly, these vanish in $d=3$, so that $\Delta_{\sigma^k} = 2k + k \gamma_{\sigma}=k\Delta_{\sigma}$, as pointed out in \cite{Leonhardt:2003du}. In other words, from a bulk point of view, the interaction energy of the $k$-particle states of the AdS scalar field appear to vanish to leading order in $1/N$ in the special case $d=3$. We will comment on this further below.

\subsection{Spinning double-trace operators}
\label{sigmadsigma}
In this section we turn our attention to the spinning operators constructed from the $\sigma$ fields, $\sigma \pl^s \sigma$ in the bosonic $O(N)$ model and the large-$N$ Gross-Neveu model. The anomalous dimensions of $\sigma \pl^s\sigma$ and many other operators in the critical vector model were computed by Lang and Ruhl in \cite{Lang:1992zw}. Analogous results for the GN model are not available to the best of our knowledge. While an extensive discussion of the GN model can be found in many places, see e.g. \cite{Derkachov:1993uw,Gracey:1992cp,Vasiliev:1992wr}, we will follow the conventions used in \cite{Manashov:2016uam} and collect below only the necessary ingredients. The momentum space propagators for $\psi$ and $\sigma$ are:
\begin{align}
D_\psi&=\frac{\slashed{p}}{p^2}\,, & D_\sigma&=\frac{\tilde{C}_{\sigma\sigma}}{(p^2)^{d/2-1+\delta}}\,, && \tilde{C}_{\sigma\sigma}=-\frac{2^d \pi ^{d/2} \Gamma (d-1)}{\Gamma \left(1-\frac{d}{2}\right) \Gamma \left(\frac{d}{2}\right)^2}\,,
\end{align}
where $\delta$ is a regulator and the only interaction vertex is $\frac{1}{\sqrt{N}}\bar{\psi}\psi\sigma$.
Let us write the full conformal dimension of $O_s\sim \sigma \pl^s\sigma$ as
\begin{align}
\label{fermionsigmadsigma}
        \Delta_{\sigma \pl^s\sigma}&=2\Delta_{\sigma}+s+\gamma_s = 2+s+2\gamma_\sigma+\gamma_s\,.
\end{align}
Note that in this and in the next subsection we adopt this definition of $\gamma_s$, which is naturally interpreted
as an interaction energy from AdS point of view.

\begin{figure}
\centering
\begin{minipage}{.5\textwidth}
  \centering
  \includegraphics[width=.6\linewidth]{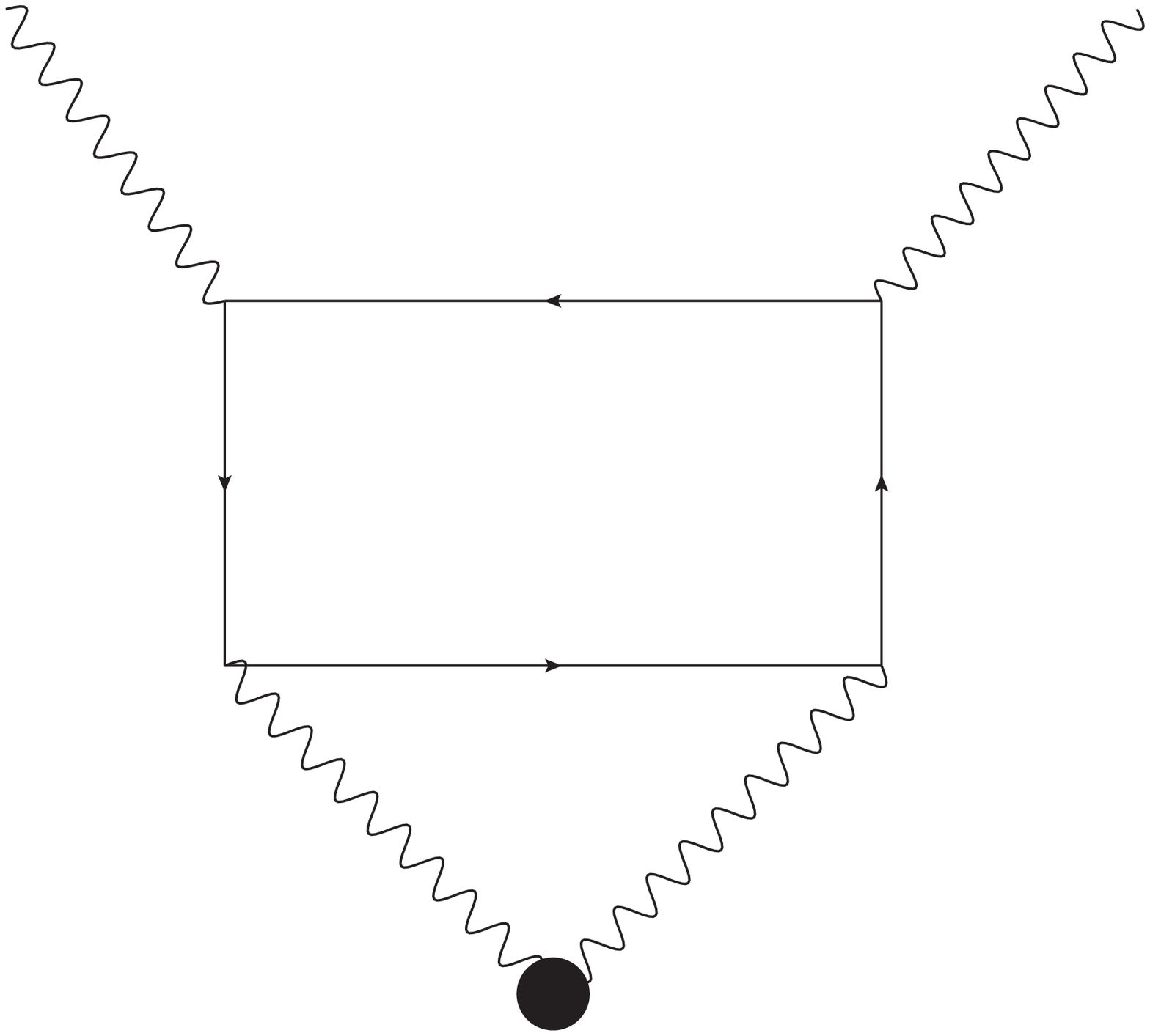}
\end{minipage}%
\begin{minipage}{.5\textwidth}
  \centering
  \includegraphics[width=.3\linewidth]{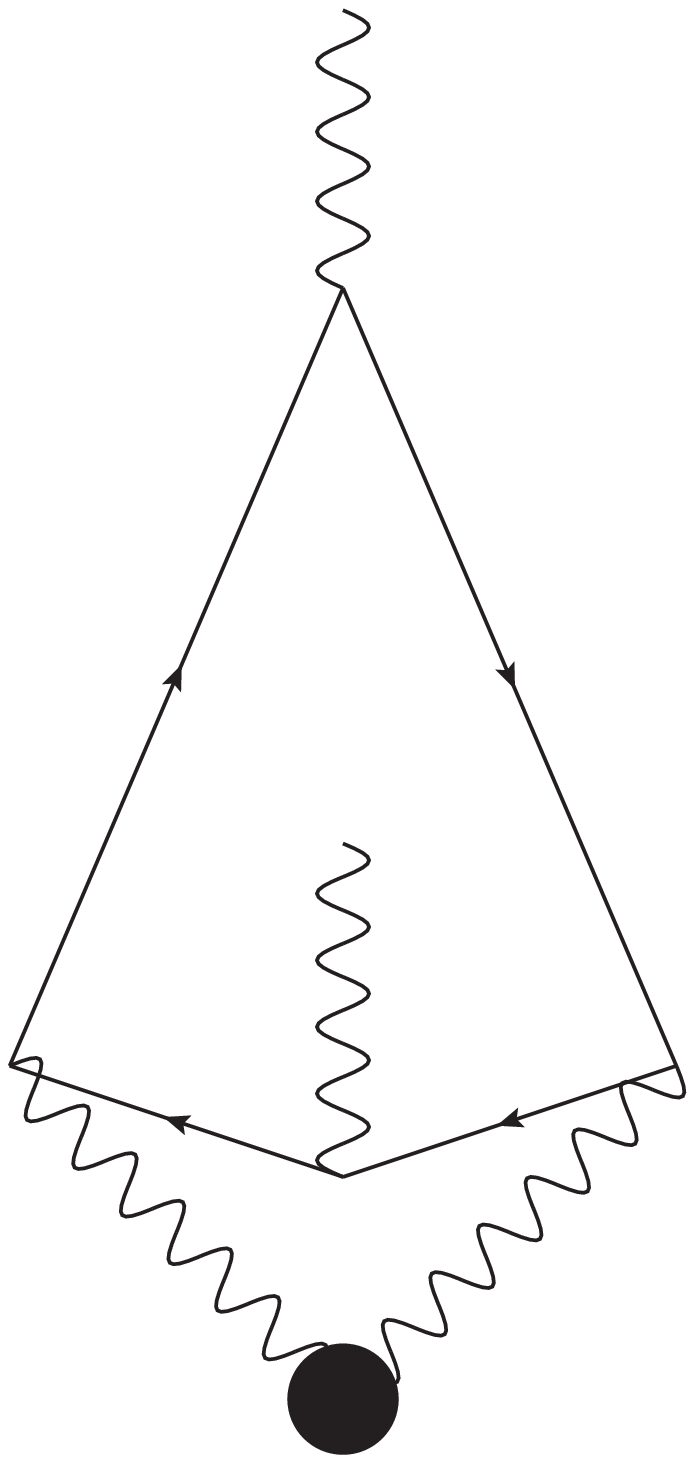}
\end{minipage}
\caption{Diagrams contributing to the anomalous dimension of $ \sigma \partial^s\sigma$ to order $1/N$.}
\label{fig:diagrams}
\end{figure}

The anomalous dimensions can be extracted by computing the 3-point functions $\langle O_s \sigma \sigma\rangle$. This is most conveniently done in momentum space, and it
is sufficient to set the momentum of the operator $O_s$ to zero. There are two diagrams, given in figure \ref{fig:diagrams}, that contribute to the $ \gamma_s$ defined in (\ref{fermionsigmadsigma}) (the leg corrections only affect $\gamma_\sigma$), where the blob corresponds to the operator insertion.
Up to $\tilde C_{\sigma\sigma}$ factors the integrals that correspond to these diagrams are, denoting by $l$ the external momentum of the $\sigma$ legs:
\begin{align}
   I_1&=\int \frac{d^{d}p}{(2\pi)^{d}}  \int \frac{d^{d}q}{(2\pi)^{d}} \frac{(z\cdot q)^s tr[\slashed p (\slashed p-\slashed q) \slashed p ( \slashed p - \slashed l)]}{(q^2)^{d-2+2\delta}(p-q)^2(p-l)^2(p^2)^2}\,  \\
   I_2&=\int \frac{d^{d}p}{(2\pi)^{d}} \int\frac{d^{d}q}{(2\pi)^{d}} \frac{(z\cdot q)^s }{(q^2)^{d-2+2\delta}}\frac{ tr[\slashed p(\slashed p+\slashed q)(\slashed p+\slashed q-\slashed l)(\slashed p-\slashed l)]}{p^2 (p-l)^2 (p+q)^2(p+q-l)^2}\,,
\end{align}
where, as usual, we conveniently take $z$ to be null, $z^2=0$. The anomalous dimensions are related to the $1/\delta$ poles of these integrals.
The second integral can be simplified by using
\begin{align}
   tr[\slashed p(\slashed p+\slashed q)(\slashed p+\slashed q-\slashed l)(\slashed p-\slashed l)]=\frac12 \left[(p-l)^2(p+q)^2+(p+q-l)^2p^2-l^2 q^2\right]
\end{align}
and dropping the last term as it does not contribute to the divergent part. The integrals are then standard and can be done with the help of
\begin{align}
\int\frac{d^{d}q}{(2\pi)^{d}} \frac{(z\cdot q)^s }{q^{2a}(q-p)^{2b}}&=\frac{ \Gamma (d/2 -b) \Gamma (a+b-d/2) \Gamma (-a+s+d/2 ) (z\cdot p)^s }{(4 \pi )^{d/2 }\Gamma (a) \Gamma (b) \Gamma (-a-b+s+d )(p^2)^{a+b-d/2 }}
\end{align}
and similar formulae. The final results are:
\begin{align}
   \gamma_{s=0}&=\frac{4(d-1) d}{d-2}\gamma_\psi\,,\\
    \label{fermionsigma}
    \gamma_{s>0}&= -2\gamma_\sigma \frac{2 d \sin \left(\frac{\pi  d}{2}\right)  \Gamma \left(\frac{d}{2}\right)^2}{\pi  (d-1)(d-2)(d-4) } \frac{\Gamma \left(s+2 -\frac{d}{2} \right)}{\Gamma \left(s+\frac{d}{2}\right)}\,.
\end{align}
Note that the $s=0$ result is not related to the $s\rightarrow 0$ limit of the non-zero spin result. This is because the residue of the $1/\delta$-pole happens to be discontinuous at $s=0$.
As far as we know, the general spin result was not derived earlier in the literature. As a check, we note that using the known anomalous dimensions
of $\sigma$ and $\sigma^2$ \cite{Gracey:1990wi}
\begin{align}
        \gamma_\sigma&=-\frac{4 (d-1)}{d-2}\gamma_\psi\,, &
        \gamma_{\sigma^2}&=4 (d-1)\gamma_\psi\,, &
        \gamma_{\sigma^2}-2\gamma_\sigma&=\frac{4 (d-1) d}{d-2}\gamma_\psi\,,
\end{align}
we find agreement between $\gamma_{s=0}$ and the last expression in the above equation. A non-trivial consistency check of the result (\ref{fermionsigma}) is
the comparison with the $4-\epsilon$ expansion of the GNY model.  Indeed, expanding (\ref{fermionsigma}) in $d=4-\epsilon$, we find
\begin{align}
   \left(  2\gamma_{\sigma}+\gamma_s  \right) \Big|_{d=4-\epsilon}&=-\frac{2 \left(3 s^2+3 s-2\right) \epsilon}{s (s+1)}+O(\epsilon^4)\,,
\end{align}
which precisely agrees with  \eqref{sigma4}.

From (\ref{fermionsigma}), we can extract the large spin behavior of the anomalous dimensions
\begin{equation}
\gamma_s = \frac{16 \sin ^2\left(\frac{\pi  d}{2}\right) \Gamma (d-2)}{\pi ^2 (4-d)}\frac{1}{s^{d-2}}+\ldots \,,
\end{equation}
suggesting that the leading term is due to the exchange of a tower of operators with twist $d-2$, i.e. the nearly conserved currents of the model at large $N$. We will comment further on this in the next subsection.

\subsubsection{Comparison to analytic bootstrap approach}
\label{sec:lcone-boots}
Let us discuss the relation of the results of the previous subsection to the general predictions \cite{Fitzpatrick:2012yx, Komargodski:2012ek,Alday:2015ewa} for the anomalous dimensions of double trace operators $O\partial^s O$, where $O$ is a scalar primary of dimension $\Delta$. One matches the anomalous dimension of the $O\partial^s O$ operators exchanged in the ``s-channel" of the four-point function $\langle OOOO\rangle$ to the ``t-channel" exchange of low-twist operators and their descendants. The expansion organizes itself as a function of conformal spin \cite{Alday:2015eya}:
\be
J^2 = (\Delta+s + \gamma_s/2) (\Delta + s + \gamma_s/2 -1)\,.
\ee
The contribution of the operator with spin $l$ and twist $\tau$ exchanged in the t-channel is then:
\be
\label{AZres}
\gamma_J (\tau,l) = -\frac{c_0(\tau,l)}{J^{\tau}}\left(   1 + \sum_{k=1}^{\infty}  \frac{c_k}{J^{2k}}    \right)\,,
\ee
where the coefficient $c_0$ controlling the leading term in the large spin expansion is known from \cite{Fitzpatrick:2012yx,Komargodski:2012ek}:
\be
\label{ZK}
c_0(\tau,l) = \frac{C^2_{OOO_{\tau}}}{C_{O_{\tau}O_{\tau}}C^2_{OO}} \frac{(-1)^l \Gamma(\tau+2l) \Gamma^2(\Delta)}{2^{l-1}\Gamma^2(\Delta-\frac{\tau}{2})\Gamma^2(l+\frac{\tau}{2})}\,.
\ee

In our case, $ O=\sigma$ and $\Delta=1+O(1/N)$, and to leading order in $1/N$ we can study the exchange of the following operators in the t-channel: $\sigma$, the tower $\sigma \pl^l \sigma$ of twist $2+O(1/N)$, and the fermionic higher-spin currents $\bar{\psi} \hat{\gamma} \deh^{l-1} \psi$ of twist $d-2+O(1/N)$. Now the contribution of the $\sigma$ operator vanishes since the three-point function $\langle \sigma \sigma \sigma \rangle = 0$ identically in the fermion model due to the parity symmetry.
Further, we note that the $\sigma \pl^l \sigma$ would contribute to (\ref{ZK}) only at the order $1/N^2$, since at tree level $\Delta=1$, $\tau =2$ and we would have hit $\Gamma^2(0)$ in the denominator, which gets resolved by expanding all the dimensions to the first order in $1/N$.

Hence, the exchange of the higher-spin currents of the form $\bar{\psi} \hat{\gamma} \deh^{l-1} \psi$ should be what reproduces the answer for the anomalous dimensions (\ref{fermionsigma}). The relevant OPE coefficients that appear in (\ref{ZK}) are explicitly computed in Appendix \ref{OPEC} as a function of the spin of the exchanged operator $l$ and are of order $1/N$. This is sufficient to reproduce the leading large spin term in the anomalous dimensions. In \cite{Alday:2015ewa}, it has been shown how to sum the contributions other the descendants in \eqref{AZres} for the exchanged operator with a twist $d-2$, which is the case at hand, and obtain the finite spin
dependence of the anomalous dimensions.  The result is
\be
\label{AZ}
\gamma_{s}(l) = -c_0(d-2,l) \frac{\Gamma(s+1)\Gamma(2\Delta+s-\frac{d}{2})}{\Gamma(s+\frac{d}{2})\Gamma(2\Delta+s-1)}\,.
\ee
Notice that in our case $\Delta=1$ is the tree-level dimension of the $\sigma$ operator, and we have to sum over the spin $\l$ of the exchanged tower. Using (\ref{ZK}) and
the coefficient $a_{cr.F,s}$ from (\ref{OPEcoef}), the sum we need to perform is:
\be
\label{F-sum}
\sum_{l=2,4,6,\ldots}^{\infty} \frac{16 (2l+d-3) \sin^2(\frac{d\pi}{2}) \Gamma^2(d-2)\Gamma(l)}{\pi^2 \Gamma(l+d-2)} = \frac{16 \sin^2(\frac{d\pi}{2})\Gamma(d-2)}{(d-4)\pi^2}\,.
\ee
Note that this sum is power-like divergent for $d<4$, and we have evaluated it by analytic continuation in $d$, as usual in dimensional regularization.
Putting it all together using (\ref{AZ}), we arrive at the same result \eqref{fermionsigma}, in a slightly different form:
\be
\label{GNsigmasigma}
\gamma_s = -\frac{16 \sin^2(\frac{d\pi}{2}) \Gamma(d-2) \Gamma(s+2-\frac{d}{2})}{N(d-4)\pi^2 \Gamma(s+\frac{d}{2})}\,.
\ee
In particular, in $d=3$ we obtain
\be
\label{GNsigmasigma3d}
\gamma_s = \frac{32}{\pi^2(2s+1)}\frac{1}{N}+O(1/N^2)\,.
\ee
Note that the anomalous dimensions are {\it positive} in $2<d<4$ for all spins. This may seem surprising, since each even spin field exchange contributes, according to (\ref{AZres})-(\ref{ZK}), a negative amount to the anomalous dimensions (from AdS point of view, we expect attractive interactions due to each even spin $l$). However, the infinite sum in (\ref{F-sum}) is divergent, and its regularization appears to yield a positive final result. The fact that this procedure gives exact agreement with the direct diagrammatic calculation of the previous subsection is a good check that the regularization of the sum makes sense.

Let us also include here some results for these operators in the scalar theory. The $1/N$ result for $\gamma_s$ has been obtained by Lang and Ruhl \cite{Lang:1992zw}, and takes the form
\be
\label{LR}
\gamma_s = -2\gamma_\sigma \frac{d(d-3)}{(d-1)(d-2)(d-4)(s+1)(s+2)}    \left(d^2-5 d+6-\frac{2 \Gamma(\frac{d}{2})  \Gamma(s-\frac{d}{2}+4)} {\Gamma(4-\frac{d}{2})\Gamma(\frac{d}{2}+s) }\right) \,,
\ee
with $\gamma_{\sigma}$ given in (\ref{gamsig-boson}).
Note that in $d=3$ both of the spin-dependent contributions vanish, and the function \eqref{LR} just reduces to $2\gamma_\sigma$, so that
$\Delta_{\sigma \partial^s \sigma}=2\Delta_{\sigma}+s$ in $d=3$. In other words, the ``binding energy" of the corresponding two-particle state
in the $AdS_4$ dual theory vanish, similarly to what observed earlier for the $\sigma^k$ operators. In the range $2<d<6$ where the scalar
CFT is unitary, we note that that the anomalous dimensions are positive in $3<d<4$, and negative in $2<d<3$ and $4<d<6$.\footnote{The negativity
in $4<d<6$ is consistent with Nachtmann theorem \cite{Nachtmann:1973mr},
since in this range the operators $\sim \sigma \partial^s\sigma$, with $\tau=2+O(1/N)$,
are the minimal twists in the $\sigma\sigma$ OPE.} In general $d$, the result (\ref{LR}) has
the large spin expansion
\be
\gamma_s =    - \frac{1}{s^2}\frac{2^d (d-3)^3 (d-2) \sin \left(\frac{\pi  d}{2}\right) \Gamma \left(\frac{d-3}{2}\right)}{N\pi ^{3/2} (d-4)  \Gamma \left(\frac{d}{2}\right)} - \frac{1}{s^{d-2}}\frac{128 (d-3) \sin^2 \left(\frac{\pi  d}{2}\right) \Gamma (d-2)}{N\pi ^2 (d-6) (d-4)^2}\,.
\label{LR-large-s}
\ee
Again, in light of the results of \cite{Fitzpatrick:2012yx, Komargodski:2012ek,Alday:2015ewa}, we clearly see two different contributions in the t-channel accounting for this behavior: the exchange of $\sigma$ operator with twist $2$, producing the $1/s^2$, and the exchange of the higher spin tower $\phi \pl^l \phi$, accounting for the $1/s^{d-2}$.
We can reproduce the full spin dependence of the contribution generated by the higher-spin tower by taking $\Delta =2$ in \eqref{AZ}
and using the bosonic OPE coefficients from \eqref{OPEcoef}:
\begin{eqnarray}
\frac{\Gamma(s-\frac{d}{2}+4)}{(s+1)(s+2)\Gamma(s+\frac{d}{2})} \sum_{l=2,4,6,\ldots}^{\infty} \frac{64 (2l+d-3) \sin^2(\frac{d\pi}{2}) \Gamma^2(d-2)\Gamma(l+1)}{\pi^2 (d-4)^2 \Gamma(l+d-3)}= \nonumber \\
= \frac{128 (d-3)\sin^2(\frac{d\pi}{2})\Gamma(d-2)}{(d-4)^2 (d-6)\pi^2} \frac{\Gamma(s-\frac{d}{2}+4)}{(s+1)(s+2)\Gamma(s+\frac{d}{2})} \,,
\end{eqnarray}
where we have again regularized the sum by analytic continuation.
This is exactly equal to the last term in the expression \eqref{LR}. The other contribution in (\ref{LR}),
which is responsible for the $1/s^2$ term in (\ref{LR-large-s}) at large spin, can be reproduced by using the known
three-point function coefficients $\langle\sigma \sigma \sigma\rangle $ \cite{Petkou:1994ad}:
\be
\frac{C^2_{\sigma\sigma\sigma}}{C^3_{\sigma\sigma}}  = \frac{1}{N}\frac{8(d-3)^2\Gamma(d-2)}{\Gamma(3-\frac{d}{2}) \Gamma^3(\frac{d}{2}-1)}\,.
\ee
Plugging this into (\ref{ZK}), we see that the resulting factor precisely agrees with the first term in (\ref{LR-large-s}). Remarkably, the full spin dependence can also
be reconstructed in this case. This is because, for the special value $\Delta=\tau=2$, it follows from the analysis of \cite{Alday:2015ewa}
that all higher order coefficients $c_k$ in (\ref{AZres}) vanish, and the finite spin answer is simply obtained from (\ref{LR-large-s}) by replacing
$1/s^2$ with $1/J^2=1/(s+1)(s+2)$.

It is interesting to compare the above calculation to the case of the free fermionic and bosonic vector models. In the free CFT, we of course
expect no anomalous dimensions for the double trace operators $\sim \bar\psi \psi  \partial^s \bar\psi \psi$ and $\sim \phi^2 \partial^s\phi^2$. But formally
we can still apply the above analytic bootstrap results, and we would then expect to recover the vanishing of the anomalous dimensions from the sum
over the infinite tower in the crossed channel. From a bulk
point of view, this would correspond to computing the tree level 4-point functions of the bulk scalar,
reading off the contribution to the binding energies of each diagram with a higher spin exchange (as
well as the quartic scalar vertex), and summing up over all spins.\footnote{Right the opposite was recently done in \cite{Bekaert:2015tva}: the quartic scalar self-interaction vertex was reconstructed in such a way that the full four-point function matches that of the free scalar CFT. } In the free fermion theory, using (\ref{ZK}) and the OPE coefficients in \ref{OPEfree},
we encounter the sum (we omit the $l$-independent overall factors, and recall that the $\bar\psi\psi$ 3-point function is zero)
\begin{equation}
\sum_{l=2,4,6,\ldots}^{\infty}\frac{(2 l+d-3) \Gamma (d+l-2)}{\Gamma (l)}=0\,,
\label{freeF-sum}
\end{equation}
and in the scalar theory
\begin{equation}
\sum_{l=0,2,4,6,\ldots}^{\infty}\frac{(2 l+d-3) \Gamma (d+l-3)}{\Gamma (l+1)}=0\,.
\end{equation}
Both of these sums, when regulated by analytic continuation in $d$, vanish. In $d=3$, one may also evaluate them by Riemann zeta-function regularization, see the next subsection. From the bulk perspective, the vanishing
of the regularized sum over ``binding energies" we encounter here is reminiscent of the vanishing of the regularized sum over one-loop vacuum energies
found in \cite{Giombi:2013fka, Giombi:2014iua}.

\subsubsection{Chern-Simons vector models in $d=3$}
\label{CSmat}
The analytic bootstrap results used in the previous subsection can be also readily applied to the bosonic and fermionic 3d Chern-Simons vector models of \cite{Giombi:2011kc,Aharony:2011jz}. We consider
$U(N_c)$ Chern-Simons theory at level $k$ coupled to a fundamental scalar or fermion, in the large $N_c$ limit with $\lambda=N_c/k$ fixed, and use the approach described above to compute the anomalous dimensions of
the spinning double trace operators $\sim \bar\phi\phi \partial^s \bar\phi\phi$ and $\sim \bar\psi\psi \partial^s \bar\psi \psi$.

Let us start with the scalar theory. The operators contributing in the $t$-channel, as before, will be the higher-spin tower (which still has twist $d-2+O(1/N_c)$, for any value of the 't Hooft coupling) and the scalar $\bar\phi \phi$. The latter has dimension $\Delta=d-2+O(1/N_c)=1+O(1/N_c)$, so its contribution can be calculated exactly as a function of spin $s$ using (\ref{AZ}). Using the results of \cite{Maldacena:2012sf} which follow from the weakly broken higher-spin symmetries, the OPE coefficients we need
still take the form \eqref{OPEfree}, dressed with factors that depends on the parameters $\tilde N$ and $\tilde\lambda$ defined in \cite{Maldacena:2012sf}, which can be fixed \cite{Aharony:2012nh, GurAri:2012is} to be
\begin{equation}
\tilde N = 2N_c\frac{\sin(\pi\lambda)}{\pi\lambda}\,,\qquad \tilde\lambda = \tan(\frac{\pi\lambda}{2})\,.
\end{equation}
Explicitly, using the results of \cite{Maldacena:2012sf}, one finds that the contribution of the higher-spin tower acquires a factor $1/\tilde{N}$, and the scalar contribution comes with a factor $\frac{1}{\tilde{N}}\frac{1}{1+\tilde{\lambda}^2}$. Putting all the factors from \eqref{AZres}, \eqref{ZK}, \eqref{AZ}, \eqref{OPEfree} together in $d=3$, we get\footnote{Keep in mind that we define
$\gamma_{\bar\phi\phi \partial^s \bar\phi\phi} = \Delta_{\bar\phi\phi \partial^s \bar\phi\phi} -s - 2\Delta_{\bar\phi\phi}$. The scalar scaling
dimension is $\Delta_{\bar\phi\phi}=1+\gamma_{\bar\phi\phi}$, but $\gamma_{\bar\phi\phi}=f_0(\lambda)/N_c+O(1/N_c^2)$ is not currently known
to all orders in $\lambda$.}
\be
\gamma_{\bar\phi\phi \partial^s \bar\phi\phi} = - \frac{2}{2s+1}\left(\frac{1}{\tilde{N}} \sum^{\infty}_{l=2,4,6,\ldots} \frac{32}{\pi^2} + \frac{1}{\tilde{N}} \frac{1}{1+\tilde{\lambda}^2} \frac{16}{\pi^2}  \right) = \frac{1}{\tilde{N}} \frac{\tilde{\lambda}^2}{1+\tilde{\lambda}^2} \frac{32}{\pi^2 (2s+1)}+O(1/N_c^2)\,,
\ee
where we used the Riemann zeta function to regulate the sum (alternatively, one can regulate the sums dimensionally as in the previous section). Note that the answer goes to zero as $\tilde{\lambda} \rightarrow 0$, as expected since in that limit we recover the free scalar CFT, and for $\tilde\lambda\rightarrow \infty$ it goes into the 3d Gross-Neveu result \eqref{GNsigmasigma3d}, as it should be in accordance with the 3d bosonization duality.\footnote{To be precise, at $\tilde\lambda\rightarrow \infty$ we get the result for
the $U(k-N_c)$ GN model, and one should keep in mind that in \eqref{GNsigmasigma3d} $N=N_c {\rm tr}1=2N_c$.}

Let us now move to the fermionic theory. The calculation is almost the same, with the difference that the scalar contribution vanishes, since the three-point function of the scalar $\bar{\psi} \psi$ vanishes. As for the higher-spin tower (which acquires a factor $1/\tilde N$ as above), after plugging all the factors in $d=3$, its contribution is proportional to the sum
\be
\sum^{\infty}_{l=2,4,6,\ldots} l^2 = 4 \zeta(-2) = 0\,
\ee
using the same zeta function regularization (equivalently, one gets the same result using the dimensionally continued sum in \eqref{freeF-sum}), and hence
we conclude that $\gamma_{\bar\psi\psi \partial^s \bar\psi\psi} = O(1/N_c^2)$ in the CS-fermion model. Note that this result is expected from
the absence of anomalous dimensions in the free fermion theory, because apart from the overall factor of $1/\tilde N$ instead of $1/N$, the sum over the higher-spin tower is
otherwise identical in the CS-fermion theory and free fermion theory, and has to vanish in the latter.  This also explains
the vanishing of the result \eqref{LR} in
the $d=3$ critical scalar model: the scalar exchange contribution is absent in this model as well ($C_{\sigma\sigma\sigma}=0$ in the 3d critical $O(N)$ model), and the higher-spin tower contribution is the same up to the overall function of $\tilde\lambda$; this is because the dimension of the external scalar is $\Delta=2$ in both cases, and the relevant OPE coefficients then coincide, see eq. (\ref{OPE-3d}). In this sense, the vanishing of the $\sigma$ binding energies in the 3d critical $O(N)$ model can be seen as a manifestation of the 3d bosonization duality.

\section*{Acknowledgments}

We would like to thank Alexander Manashov, Eric Perlmutter, David Poland and Sasha Zhiboedov for useful discussions and correspondence.
The work of SG and VK was supported in part by the US NSF under Grant No.~PHY-1620542.
The work of ES was supported  in part  by the Russian Science Foundation grant 14-42-00047 in association with Lebedev Physical Institute and by the DFG Transregional Collaborative Research Centre TRR 33 and the DFG cluster of excellence ``Origin and Structure of the Universe". ES, VK and SG would like to thank Munich Institute for Astro- and Particle Physics (MIAPP) of the DFG cluster of excellence ``Origin and Structure of the Universe" for the hospitality.  VK and SG also thank the GGI Institute for the hospitality during completion of this work.

\begin{appendix}
\renewcommand{\thesection}{\Alph{section}}
\renewcommand{\theequation}{\Alph{section}.\arabic{equation}}
\setcounter{equation}{0}\setcounter{section}{0}

\section{OPE Coefficients from AdS/CFT}
\label{OPEC}
AdS/CFT relates the OPE coefficients of the UV and IR duals, i.e. the duals of the same bulk theory for different choice of boundary conditions within the unitarity window \cite{Klebanov:1999tb,Klebanov:2002ja}. We can use this fact to compute some of the OPE coefficients in the critical $O(N)$ vector model and Gross-Neveu model using the OPE coefficients of the free scalar and free fermion, respectively.

The OPE coefficients we are interested in correspond to three-point function $\langle J_s J_0 J_0\rangle$, where $J_0$ has scaling dimension $\Delta=2$ for the critical model and $\Delta=1$ for GN, and $J_s$ are the totally symmetric (nearly) conserved currents with twist $\tau=d-2+O(1/N)$.
We define the normalized OPE coefficients $a_{s}$ as
\begin{align}
a_s &\equiv \frac{C_{s00}^2}{C_{ss} C_{00}^2}\label{HSinvariant}\,,
\end{align}
where we the coefficients on the right-hand side are defined in our conventions by
\begin{equation}
\begin{aligned}
&\langle J_0(x_1)J_0(x_2)\rangle = \frac{C_{00}}{x_{12}^{2\Delta}}\,,\qquad
\langle \hat{J}_s (x_1,z_1) \hat{J}_s (x_2,z_2) =C_{ss} \frac{\left(z_1\cdot z_2
-\frac{2z_1\cdot x_{12} z_2 \cdot x_{12}}{x_{12}^2}\right)^s}{x^{2\Delta_s}_{12}}\,, \\
&\langle \hat{J}_s(x_1,z_1) J_0(x_2) J_0(x_3) \rangle  = C_{s00}
\frac{\left(\frac{z_1\cdot x_{13}}{x^2_{13}}-\frac{z_1\cdot x_{12}}{x^2_{12}}\right)^s}{x^{\tau}_{12}x^{\tau}_{13}x^{2\Delta - \tau}_{23}}
\equiv C_{s00} \JOOst{s}{\Delta}\,,
\end{aligned}
\end{equation}
where $z_1$, $z_2$ are null vectors.
The OPE coefficients for the free fermion theory were given in (\ref{Cs00-fer}), and for the scalar can be found in \cite{Diaz:2006nm,Bekaert:2015tva, Skvortsov:2015pea, Sleight:2016dba},
and we obtain:
\begin{align}
\label{OPEfree}
a_{B,s}&=\frac{\sqrt{\pi } 2^{-d-s+7} \Gamma \left(\frac{d}{2}+s-1\right) \Gamma (d+s-3)}{N \Gamma \left(\frac{d}{2}-1\right)^2 \Gamma (s+1) \Gamma \left(\frac{d-3}{2}+s\right)}\,,\\
a_{F,s}&=\frac{\sqrt{\pi } (-)^{s} 2^{-d-s+5} \Gamma \left(\frac{d}{2}+s-1\right) \Gamma (d+s-2)}{N \Gamma \left(\frac{d}{2}\right)^2 \Gamma (s) \Gamma \left(\frac{d-3}{2}+s\right)}\,.
\end{align}

The three-point function of one higher-spin current $J_s$ and two scalar operators of dimension $\Delta$ comes from the unique cubic interaction vertex in $AdS$
\begin{align}
g_{s}\int \Fron_{\mm(s)} &\nabla^{\mm(s)}\Fron_0 \Fron_0= g_{s}\tilde{b}_{s}\times \JOOst{s}{\Delta}\,,\\
\tilde{b}_{s}&=\frac{2^{-5+2 s} \pi ^{-d/2} (-3+d+2 s)\Gamma\left[-1+\frac{d}{2}+s\right]^3\Gamma[-3+d+s]\Gamma[-1+s+\Delta ]^2}{\Gamma[-2+d+2 s]^2\Gamma[\Delta ]^2}\,,
\end{align}
where $g_s$ is the coupling constant of the bulk theory and $b_s$ is a nontrivial factor produced by integrating the vertex on boundary-to-bulk propagators \cite{Costa:2014kfa}. The coupling constant $g_s$ can be chosen as to reproduce the OPE coefficients $C_{s00}$ for free scalar $\Delta=d-2$ or free fermion $\Delta=d-1$. The results \cite{Bekaert:2015tva,Skvortsov:2015pea} are
\begin{align}
\text{boson}&: && g^B_s=\frac{1}{\sqrt{N}} \frac{\pi ^{\frac{d-3}{4}} 2^{\frac{1}{2} (3 d+s-1)} \Gamma \left(\frac{d-1}{2}\right) }{\Gamma (d+s-3)  }\sqrt{\frac{\Gamma \left(\frac{d-1}{2}+s\right)}{\Gamma (s+1)}}\,,\\
\text{fermion}&: && (g^F_s)^2=(g^B_s)^2 \frac{s}{(d+s-3)}\,.
\end{align}
Next, we can change the boundary conditions to $\Delta=2$ and $\Delta=1$, respectively, and recompute the bulk integral, i.e. $b_s$. The result should give the OPE coefficients  for the critical models:
\begin{equation}
\begin{aligned}
\label{OPEcoef}
a_{cr.B,s}&=\frac{2^{d-s+1} \Gamma \left(\frac{d-1}{2}\right)^2 \Gamma (s+1) \Gamma \left(\frac{d}{2}+s-1\right)}{\sqrt{\pi } N \Gamma \left(\frac{d-3}{2}+s\right) \Gamma (d+s-3)}\,,\\
a_{cr.F,s}&=\frac{ 2^{d-s+1} \Gamma \left(\frac{d-1}{2}\right)^2 \Gamma (s) \Gamma \left(\frac{d}{2}+s-1\right)}{\sqrt{\pi } N \Gamma \left(\frac{d-3}{2}+s\right) \Gamma (d+s-2)}\,.
\end{aligned}
\end{equation}
The same result can, of course, be obtained on the CFT side by attaching two propagators of the $\sigma$-field to the three-point functions of the free theories. As
explained in \cite{Hartman:2006dy, Giombi:2011ya}, the procedure of attaching a $\sigma$ line on the CFT side is in one-to-one correspondence with changing the boundary condition on the bulk scalar propagator, and hence one is essentially guaranteed to obtain the same result. Nevertheless, to double-check our results we have explicitly computed (\ref{OPEcoef}) directly on the CFT side, and obtained the same result.

Note that in $d=3$, when the leading large $N$ dimensions of the scalar operator coincide in free fermion/critical scalar and free scalar/critical fermion, we have
\begin{equation}
\begin{aligned}
&a_{F,s}|_{d=3} = a_{cr.B,s}|_{d=3}\,,\\
&a_{B,s}|_{d=3} = a_{cr.F,s}|_{d=3}\,.
\label{OPE-3d}
\end{aligned}
\end{equation}

\end{appendix}

\begingroup\raggedright\endgroup


\begin{thebibliography}{10}

\bibitem{Gross:1974jv}
D.~J. Gross and A.~Neveu, ``{Dynamical Symmetry Breaking in Asymptotically Free
  Field Theories},'' {\em Phys.Rev.} {\bf D10} (1974) 3235.

\bibitem{Wilson:1972cf}
K.~G. Wilson, ``{Quantum field theory models in less than four-dimensions},''
  {\em Phys. Rev.} {\bf D7} (1973) 2911--2926.

\bibitem{Gross:1975vu}
D.~J. Gross, ``{Applications of the Renormalization Group to High-Energy
  Physics},'' {\em Conf. Proc.} {\bf C7507281} (1975) 141--250.

\bibitem{Moshe:2003xn}
M.~Moshe and J.~Zinn-Justin, ``{Quantum field theory in the large N limit: A
  Review},'' {\em Phys. Rept.} {\bf 385} (2003) 69--228,
  \href{http://xxx.lanl.gov/abs/hep-th/0306133}{{\tt hep-th/0306133}}.

\bibitem{ZinnJustin:1991yn}
J.~Zinn-Justin, ``{Four fermion interaction near four-dimensions},'' {\em Nucl.
  Phys.} {\bf B367} (1991) 105--122.

\bibitem{Hasenfratz:1991it}
A.~Hasenfratz, P.~Hasenfratz, K.~Jansen, J.~Kuti, and Y.~Shen, ``{The
  Equivalence of the top quark condensate and the elementary Higgs field},''
  {\em Nucl. Phys.} {\bf B365} (1991) 79--97.

\bibitem{Karkkainen:1993ef}
L.~Karkkainen, R.~Lacaze, P.~Lacock, and B.~Petersson, ``{Critical behavior of
  the 3-d Gross-Neveu and Higgs-Yukawa models},'' {\em Nucl. Phys.} {\bf B415}
  (1994) 781--796, \href{http://xxx.lanl.gov/abs/hep-lat/9310020}{{\tt
  hep-lat/9310020}}. [Erratum: Nucl. Phys.B438,650(1995)].

\bibitem{Fei:2014yja}
L.~Fei, S.~Giombi, and I.~R. Klebanov, ``{Critical $O(N)$ Models in
  $6-\epsilon$ Dimensions},'' {\em Phys.Rev.} {\bf D90} (2014) 025018,
  \href{http://xxx.lanl.gov/abs/1404.1094}{{\tt 1404.1094}}.

\bibitem{Diab:2016spb}
K.~Diab, L.~Fei, S.~Giombi, I.~R. Klebanov, and G.~Tarnopolsky, ``{On $C_J$ and
  $C_T$ in the Gross-Neveu and $O(N)$ Models},'' {\em J. Phys.} {\bf A49}
  (2016), no.~40 405402, \href{http://xxx.lanl.gov/abs/1601.07198}{{\tt
  1601.07198}}.

\bibitem{Fei:2016sgs}
L.~Fei, S.~Giombi, I.~R. Klebanov, and G.~Tarnopolsky, ``{Yukawa CFTs and
  Emergent Supersymmetry},'' {\em PTEP} {\bf 2016} (2016), no.~12 12C105,
  \href{http://xxx.lanl.gov/abs/1607.05316}{{\tt 1607.05316}}.

\bibitem{Anselmi:1998ms}
D.~Anselmi, ``{The N=4 quantum conformal algebra},'' {\em Nucl. Phys.} {\bf
  B541} (1999) 369--385, \href{http://xxx.lanl.gov/abs/hep-th/9809192}{{\tt
  hep-th/9809192}}.

\bibitem{Belitsky:2007jp}
A.~V. Belitsky, J.~Henn, C.~Jarczak, D.~Mueller, and E.~Sokatchev, ``{Anomalous
  dimensions of leading twist conformal operators},'' {\em Phys. Rev.} {\bf
  D77} (2008) 045029, \href{http://xxx.lanl.gov/abs/0707.2936}{{\tt
  0707.2936}}.

\bibitem{Skvortsov:2015pea}
E.~D. Skvortsov, ``{On (Un)Broken Higher-Spin Symmetry in Vector Models},'' in
  {\em {Proceedings, International Workshop on Higher Spin Gauge Theories:
  Singapore, Singapore, November 4-6, 2015}}, pp.~103--137, 2017.
\newblock \href{http://xxx.lanl.gov/abs/1512.05994}{{\tt 1512.05994}}.

\bibitem{Giombi:2016hkj}
S.~Giombi and V.~Kirilin, ``{Anomalous dimensions in CFT with weakly broken
  higher spin symmetry},'' {\em JHEP} {\bf 11} (2016) 068,
  \href{http://xxx.lanl.gov/abs/1601.01310}{{\tt 1601.01310}}.

\bibitem{Giombi:2016zwa}
S.~Giombi, V.~Gurucharan, V.~Kirilin, S.~Prakash, and E.~Skvortsov, ``{On the
  Higher-Spin Spectrum in Large N Chern-Simons Vector Models},''
  \href{http://xxx.lanl.gov/abs/1610.08472}{{\tt 1610.08472}}.

\bibitem{Giombi:2011kc}
S.~Giombi, S.~Minwalla, S.~Prakash, S.~P. Trivedi, S.~R. Wadia, {\em et.~al.},
  ``{Chern-Simons Theory with Vector Fermion Matter},'' {\em Eur.Phys.J.} {\bf
  C72} (2012) 2112, \href{http://xxx.lanl.gov/abs/1110.4386}{{\tt 1110.4386}}.

\bibitem{Aharony:2011jz}
O.~Aharony, G.~Gur-Ari, and R.~Yacoby, ``{d=3 Bosonic Vector Models Coupled to
  Chern-Simons Gauge Theories},'' {\em JHEP} {\bf 1203} (2012) 037,
  \href{http://xxx.lanl.gov/abs/1110.4382}{{\tt 1110.4382}}.

\bibitem{Muta:1976js}
T.~Muta and D.~S. Popovic, ``{Anomalous Dimensions of Composite Operators in
  the Gross-Neveu Model in Two + Epsilon Dimensions},'' {\em Prog. Theor.
  Phys.} {\bf 57} (1977) 1705.

\bibitem{Manashov:2016uam}
A.~N. Manashov and E.~D. Skvortsov, ``{Higher-spin currents in the Gross-Neveu
  model at $1/n^2$},'' \href{http://xxx.lanl.gov/abs/1610.06938}{{\tt
  1610.06938}}.

\bibitem{Vasiliev:1992av}
M.~A. Vasiliev, ``{More on equations of motion for interacting massless fields
  of all spins in (3+1)-dimensions},'' {\em Phys.Lett.} {\bf B285} (1992)
  225--234.

\bibitem{Giombi:2012ms}
S.~Giombi and X.~Yin, ``{The Higher Spin/Vector Model Duality},'' {\em J.Phys.}
  {\bf A46} (2013) 214003, \href{http://xxx.lanl.gov/abs/1208.4036}{{\tt
  1208.4036}}.

\bibitem{Didenko:2014dwa}
V.~E. Didenko and E.~D. Skvortsov, ``{Elements of Vasiliev theory},''
  \href{http://xxx.lanl.gov/abs/1401.2975}{{\tt 1401.2975}}.

\bibitem{Giombi:2016ejx}
S.~Giombi, ``{TASI Lectures on the Higher Spin - CFT duality},''
  \href{http://xxx.lanl.gov/abs/1607.02967}{{\tt 1607.02967}}.

\bibitem{Vasiliev:2003ev}
M.~Vasiliev, ``{Nonlinear equations for symmetric massless higher spin fields
  in (A)dS(d)},'' {\em Phys.Lett.} {\bf B567} (2003) 139--151,
  \href{http://xxx.lanl.gov/abs/hep-th/0304049}{{\tt hep-th/0304049}}.

\bibitem{Klebanov:1999tb}
I.~R. Klebanov and E.~Witten, ``{AdS / CFT correspondence and symmetry
  breaking},'' {\em Nucl. Phys.} {\bf B556} (1999) 89--114,
  \href{http://xxx.lanl.gov/abs/hep-th/9905104}{{\tt hep-th/9905104}}.

\bibitem{Klebanov:2002ja}
I.~Klebanov and A.~Polyakov, ``{AdS dual of the critical O(N) vector model},''
  {\em Phys.Lett.} {\bf B550} (2002) 213--219,
  \href{http://xxx.lanl.gov/abs/hep-th/0210114}{{\tt hep-th/0210114}}.

\bibitem{Girardello:2002pp}
L.~Girardello, M.~Porrati, and A.~Zaffaroni, ``{3-D interacting CFTs and
  generalized Higgs phenomenon in higher spin theories on AdS},'' {\em Phys.
  Lett.} {\bf B561} (2003) 289--293,
  \href{http://xxx.lanl.gov/abs/hep-th/0212181}{{\tt hep-th/0212181}}.

\bibitem{Rychkov:2015naa}
S.~Rychkov and Z.~M. Tan, ``{The $\epsilon$-expansion from conformal field
  theory},'' {\em J. Phys.} {\bf A48} (2015), no.~29 29FT01,
  \href{http://xxx.lanl.gov/abs/1505.00963}{{\tt 1505.00963}}.

\bibitem{Basu:2015gpa}
P.~Basu and C.~Krishnan, ``{$\epsilon$-expansions near three dimensions from
  conformal field theory},'' {\em JHEP} {\bf 11} (2015) 040,
  \href{http://xxx.lanl.gov/abs/1506.06616}{{\tt 1506.06616}}.

\bibitem{Sen:2015doa}
K.~Sen and A.~Sinha, ``{On critical exponents without Feynman diagrams},'' {\em
  J. Phys.} {\bf A49} (2016), no.~44 445401,
  \href{http://xxx.lanl.gov/abs/1510.07770}{{\tt 1510.07770}}.

\bibitem{Ghosh:2015opa}
S.~Ghosh, R.~K. Gupta, K.~Jaswin, and A.~A. Nizami, ``{$\epsilon$-Expansion in
  the Gross-Neveu model from conformal field theory},'' {\em JHEP} {\bf 03}
  (2016) 174, \href{http://xxx.lanl.gov/abs/1510.04887}{{\tt 1510.04887}}.

\bibitem{Raju:2015fza}
A.~Raju, ``{$\epsilon$-Expansion in the Gross-Neveu CFT},''
  \href{http://xxx.lanl.gov/abs/1510.05287}{{\tt 1510.05287}}.

\bibitem{Manashov:2015fha}
A.~N. Manashov and M.~Strohmaier, ``{Conformal constraints for anomalous
  dimensions of leading twist operators},'' {\em Eur. Phys. J.} {\bf C75}
  (2015), no.~8 363, \href{http://xxx.lanl.gov/abs/1503.04670}{{\tt
  1503.04670}}.

\bibitem{Bashmakov:2016uqk}
V.~Bashmakov, M.~Bertolini, and H.~Raj, ``{Broken current anomalous dimensions,
  conformal manifolds and RG flows},''
  \href{http://xxx.lanl.gov/abs/1609.09820}{{\tt 1609.09820}}.

\bibitem{Bashmakov:2016pcg}
V.~Bashmakov, M.~Bertolini, L.~Di~Pietro, and H.~Raj, ``{Scalar Multiplet
  Recombination at Large N and Holography},'' {\em JHEP} {\bf 05} (2016) 183,
  \href{http://xxx.lanl.gov/abs/1603.00387}{{\tt 1603.00387}}.

\bibitem{Nii:2016lpa}
K.~Nii, ``{Classical equation of motion and Anomalous dimensions at leading
  order},'' {\em JHEP} {\bf 07} (2016) 107,
  \href{http://xxx.lanl.gov/abs/1605.08868}{{\tt 1605.08868}}.

\bibitem{Roumpedakis:2016qcg}
K.~Roumpedakis, ``{Leading Order Anomalous Dimensions at the Wilson-Fisher
  Fixed Point from CFT},'' \href{http://xxx.lanl.gov/abs/1612.08115}{{\tt
  1612.08115}}.

\bibitem{Liendo:2017wsn}
P.~Liendo, ``{Revisiting the dilatation operator of the Wilson-Fisher
  fixed-point},'' \href{http://xxx.lanl.gov/abs/1701.04830}{{\tt 1701.04830}}.

\bibitem{Vasiliev:1982dc}
A.~Vasiliev, Y.~Pismak, and Y.~Khonkonen, ``{$1/N$ Expansion: Calculation of
  the Exponent $\eta$ in the Order 1/$N^3$ by the Conformal Bootstrap
  Method},'' {\em Theor.Math.Phys.} {\bf 50} (1982) 127--134.

\bibitem{Vasiliev:1992wr}
A.~N. Vasiliev, S.~E. Derkachov, N.~A. Kivel, and A.~S. Stepanenko, ``{The 1/n
  expansion in the Gross-Neveu model: Conformal bootstrap calculation of the
  index eta in order 1/n**3},'' {\em Theor. Math. Phys.} {\bf 94} (1993)
  127--136. [Teor. Mat. Fiz.94,179(1993)].

\bibitem{Gracey:1990wi}
J.~A. Gracey, ``{Calculation of exponent eta to O(1/N**2) in the O(N)
  Gross-Neveu model},'' {\em Int. J. Mod. Phys.} {\bf A6} (1991) 395--408.
  [Erratum: Int. J. Mod. Phys.A6,2755(1991)].

\bibitem{Nachtmann:1973mr}
O.~Nachtmann, ``{Positivity constraints for anomalous dimensions},'' {\em Nucl.
  Phys.} {\bf B63} (1973) 237--247.

\bibitem{Lang:1992zw}
K.~Lang and W.~Ruhl, ``{The Critical O(N) sigma model at dimensions 2 < d < 4:
  Fusion coefficients and anomalous dimensions},'' {\em Nucl. Phys.} {\bf B400}
  (1993) 597--623.

\bibitem{Fitzpatrick:2012yx}
A.~L. Fitzpatrick, J.~Kaplan, D.~Poland, and D.~Simmons-Duffin, ``{The Analytic
  Bootstrap and AdS Superhorizon Locality},'' {\em JHEP} {\bf 1312} (2013) 004,
  \href{http://xxx.lanl.gov/abs/1212.3616}{{\tt 1212.3616}}.

\bibitem{Komargodski:2012ek}
Z.~Komargodski and A.~Zhiboedov, ``{Convexity and Liberation at Large Spin},''
  {\em JHEP} {\bf 11} (2013) 140, \href{http://xxx.lanl.gov/abs/1212.4103}{{\tt
  1212.4103}}.

\bibitem{Kaviraj:2015cxa}
A.~Kaviraj, K.~Sen, and A.~Sinha, ``{Analytic bootstrap at large spin},'' {\em
  JHEP} {\bf 11} (2015) 083, \href{http://xxx.lanl.gov/abs/1502.01437}{{\tt
  1502.01437}}.

\bibitem{Alday:2015eya}
L.~F. Alday, A.~Bissi, and T.~Lukowski, ``{Large spin systematics in CFT},''
  {\em JHEP} {\bf 11} (2015) 101,
  \href{http://xxx.lanl.gov/abs/1502.07707}{{\tt 1502.07707}}.

\bibitem{Dey:2016zbg}
P.~Dey, A.~Kaviraj, and K.~Sen, ``{More on analytic bootstrap for O(N)
  models},'' {\em JHEP} {\bf 06} (2016) 136,
  \href{http://xxx.lanl.gov/abs/1602.04928}{{\tt 1602.04928}}.

\bibitem{Kaviraj:2015xsa}
A.~Kaviraj, K.~Sen, and A.~Sinha, ``{Universal anomalous dimensions at large
  spin and large twist},'' {\em JHEP} {\bf 07} (2015) 026,
  \href{http://xxx.lanl.gov/abs/1504.00772}{{\tt 1504.00772}}.

\bibitem{Li:2015rfa}
D.~Li, D.~Meltzer, and D.~Poland, ``{Non-Abelian Binding Energies from the
  Lightcone Bootstrap},'' {\em JHEP} {\bf 02} (2016) 149,
  \href{http://xxx.lanl.gov/abs/1510.07044}{{\tt 1510.07044}}.

\bibitem{Alday:2015ota}
L.~F. Alday and A.~Zhiboedov, ``{Conformal Bootstrap With Slightly Broken
  Higher Spin Symmetry},'' {\em JHEP} {\bf 06} (2016) 091,
  \href{http://xxx.lanl.gov/abs/1506.04659}{{\tt 1506.04659}}.

\bibitem{Alday:2015ewa}
L.~F. Alday and A.~Zhiboedov, ``{An Algebraic Approach to the Analytic
  Bootstrap},'' \href{http://xxx.lanl.gov/abs/1510.08091}{{\tt 1510.08091}}.

\bibitem{Alday:2016njk}
L.~F. Alday, ``{Large Spin Perturbation Theory},''
  \href{http://xxx.lanl.gov/abs/1611.01500}{{\tt 1611.01500}}.

\bibitem{Alday:2016jfr}
L.~F. Alday, ``{Solving CFTs with Weakly Broken Higher Spin Symmetry},''
  \href{http://xxx.lanl.gov/abs/1612.00696}{{\tt 1612.00696}}.

\bibitem{Simmons-Duffin:2016wlq}
D.~Simmons-Duffin, ``{The Lightcone Bootstrap and the Spectrum of the 3d Ising
  CFT},'' \href{http://xxx.lanl.gov/abs/1612.08471}{{\tt 1612.08471}}.

\bibitem{Leonhardt:2003du}
T.~Leonhardt and W.~Ruhl, ``{The Minimal conformal O(N) vector sigma model at d
  = 3},'' {\em J. Phys.} {\bf A37} (2004) 1403--1413,
  \href{http://xxx.lanl.gov/abs/hep-th/0308111}{{\tt hep-th/0308111}}.

\bibitem{Maldacena:2012sf}
J.~Maldacena and A.~Zhiboedov, ``{Constraining conformal field theories with a
  slightly broken higher spin symmetry},'' {\em Class. Quant. Grav.} {\bf 30}
  (2013) 104003, \href{http://xxx.lanl.gov/abs/1204.3882}{{\tt 1204.3882}}.

\bibitem{Aharony:2012nh}
O.~Aharony, G.~Gur-Ari, and R.~Yacoby, ``{Correlation Functions of Large N
  Chern-Simons-Matter Theories and Bosonization in Three Dimensions},'' {\em
  JHEP} {\bf 1212} (2012) 028, \href{http://xxx.lanl.gov/abs/1207.4593}{{\tt
  1207.4593}}.

\bibitem{Vasiliev:2004cm}
M.~A. Vasiliev, ``{Higher spin superalgebras in any dimension and their
  representations},'' {\em JHEP} {\bf 12} (2004) 046,
  \href{http://xxx.lanl.gov/abs/hep-th/0404124}{{\tt hep-th/0404124}}.

\bibitem{Alkalaev:2012rg}
K.~Alkalaev, ``{Mixed-symmetry tensor conserved currents and AdS/CFT
  correspondence},'' {\em J. Phys.} {\bf A46} (2013) 214007,
  \href{http://xxx.lanl.gov/abs/1207.1079}{{\tt 1207.1079}}.

\bibitem{Dobrev:1975ru}
V.~K. Dobrev, V.~B. Petkova, S.~G. Petrova, and I.~T. Todorov, ``{Dynamical
  Derivation of Vacuum Operator Product Expansion in Euclidean Conformal
  Quantum Field Theory},'' {\em Phys. Rev.} {\bf D13} (1976) 887.

\bibitem{Craigie:1983fb}
N.~S. Craigie, V.~K. Dobrev, and I.~T. Todorov, ``{Conformally Covariant
  Composite Operators in Quantum Chromodynamics},'' {\em Annals Phys.} {\bf
  159} (1985) 411--444.

\bibitem{Costa:2011mg}
M.~S. Costa, J.~Penedones, D.~Poland, and S.~Rychkov, ``{Spinning Conformal
  Correlators},'' {\em JHEP} {\bf 11} (2011) 071,
  \href{http://xxx.lanl.gov/abs/1107.3554}{{\tt 1107.3554}}.

\bibitem{Gracey:1992cp}
J.~A. Gracey, ``{Anomalous mass dimension at O(1/N**2) in the O(N) Gross-Neveu
  model},'' {\em Phys. Lett.} {\bf B297} (1992) 293--297.

\bibitem{Gracey:1993kc}
J.~A. Gracey, ``{Computation of critical exponent eta at O(1/N**3) in the four
  Fermi model in arbitrary dimensions},'' {\em Int. J. Mod. Phys.} {\bf A9}
  (1994) 727--744, \href{http://xxx.lanl.gov/abs/hep-th/9306107}{{\tt
  hep-th/9306107}}.

\bibitem{MZ}
J.~Maldacena and A.~Zhiboedov, ``{Constraining conformal field theories with a
  slightly broken higher spin symmetry},'' {\em Class.Quant.Grav.} {\bf 30}
  (2013) 104003, \href{http://xxx.lanl.gov/abs/1204.3882}{{\tt 1204.3882}}.

\bibitem{Hikida:2016cla}
Y.~Hikida and T.~Wada, ``{Anomalous dimensions of higher spin currents in large
  N CFTs},'' {\em JHEP} {\bf 01} (2017) 032,
  \href{http://xxx.lanl.gov/abs/1610.05878}{{\tt 1610.05878}}.

\bibitem{Alday:2007mf}
L.~F. Alday and J.~M. Maldacena, ``{Comments on operators with large spin},''
  {\em JHEP} {\bf 11} (2007) 019, \href{http://xxx.lanl.gov/abs/0708.0672}{{\tt
  0708.0672}}.

\bibitem{Thomas}
S.~Thomas, ``{Emergent Supersymmetry},'' {\em Seminar at KITP} (2005).

\bibitem{Grover:2013rc}
T.~Grover, D.~N. Sheng, and A.~Vishwanath, ``{Emergent Space-Time Supersymmetry
  at the Boundary of a Topological Phase},'' {\em Science} {\bf 344} (2014),
  no.~6181 280--283, \href{http://xxx.lanl.gov/abs/1301.7449}{{\tt 1301.7449}}.

\bibitem{Bashkirov:2013vya}
D.~Bashkirov, ``{Bootstrapping the N=1 SCFT in three dimensions},''
  \href{http://xxx.lanl.gov/abs/1310.8255}{{\tt 1310.8255}}.

\bibitem{Shimada:2015gda}
H.~Shimada and S.~Hikami, ``{Fractal dimensions of self-avoiding walks and
  Ising high-temperature graphs in 3D conformal bootstrap},''
  \href{http://xxx.lanl.gov/abs/1509.04039}{{\tt 1509.04039}}.

\bibitem{Iliesiu:2015qra}
L.~Iliesiu, F.~Kos, D.~Poland, S.~S. Pufu, D.~Simmons-Duffin, and R.~Yacoby,
  ``{Bootstrapping 3D Fermions},'' {\em JHEP} {\bf 03} (2016) 120,
  \href{http://xxx.lanl.gov/abs/1508.00012}{{\tt 1508.00012}}.

\bibitem{Petkou:1994ad}
A.~Petkou, ``{Conserved currents, consistency relations and operator product
  expansions in the conformally invariant O(N) vector model},'' {\em Annals
  Phys.} {\bf 249} (1996) 180--221,
  \href{http://xxx.lanl.gov/abs/hep-th/9410093}{{\tt hep-th/9410093}}.

\bibitem{Derkachov:1993uw}
S.~E. Derkachov, N.~A. Kivel, A.~S. Stepanenko, and A.~N. Vasiliev, ``{On
  calculation in 1/n expansions of critical exponents in the Gross-Neveu model
  with the conformal technique},''
  \href{http://xxx.lanl.gov/abs/hep-th/9302034}{{\tt hep-th/9302034}}.

\bibitem{Bekaert:2015tva}
X.~Bekaert, J.~Erdmenger, D.~Ponomarev, and C.~Sleight, ``{Quartic AdS
  Interactions in Higher-Spin Gravity from Conformal Field Theory},'' {\em
  JHEP} {\bf 11} (2015) 149, \href{http://xxx.lanl.gov/abs/1508.04292}{{\tt
  1508.04292}}.

\bibitem{Giombi:2013fka}
S.~Giombi and I.~R. Klebanov, ``{One Loop Tests of Higher Spin AdS/CFT},'' {\em
  JHEP} {\bf 12} (2013) 068, \href{http://xxx.lanl.gov/abs/1308.2337}{{\tt
  1308.2337}}.

\bibitem{Giombi:2014iua}
S.~Giombi, I.~R. Klebanov, and B.~R. Safdi, ``{Higher Spin AdS$_{d+1}$/CFT$_d$
  at One Loop},'' {\em Phys. Rev.} {\bf D89} (2014), no.~8 084004,
  \href{http://xxx.lanl.gov/abs/1401.0825}{{\tt 1401.0825}}.

\bibitem{GurAri:2012is}
G.~Gur-Ari and R.~Yacoby, ``{Correlators of Large N Fermionic Chern-Simons
  Vector Models},'' {\em JHEP} {\bf 1302} (2013) 150,
  \href{http://xxx.lanl.gov/abs/1211.1866}{{\tt 1211.1866}}.

\bibitem{Diaz:2006nm}
D.~E. Diaz and H.~Dorn, ``{On the AdS higher spin / O(N) vector model
  correspondence: Degeneracy of the holographic image},'' {\em JHEP} {\bf 07}
  (2006) 022, \href{http://xxx.lanl.gov/abs/hep-th/0603084}{{\tt
  hep-th/0603084}}.

\bibitem{Sleight:2016dba}
C.~Sleight and M.~Taronna, ``{Higher Spin Interactions from Conformal Field
  Theory: The Complete Cubic Couplings},'' {\em Phys. Rev. Lett.} {\bf 116}
  (2016), no.~18 181602, \href{http://xxx.lanl.gov/abs/1603.00022}{{\tt
  1603.00022}}.

\bibitem{Costa:2014kfa}
M.~S. Costa, V.~Gonçalves, and J.~Penedones, ``{Spinning AdS Propagators},''
  {\em JHEP} {\bf 09} (2014) 064, \href{http://xxx.lanl.gov/abs/1404.5625}{{\tt
  1404.5625}}.

\bibitem{Hartman:2006dy}
T.~Hartman and L.~Rastelli, ``{Double-trace deformations, mixed boundary
  conditions and functional determinants in AdS/CFT},'' {\em JHEP} {\bf 01}
  (2008) 019, \href{http://xxx.lanl.gov/abs/hep-th/0602106}{{\tt
  hep-th/0602106}}.

\bibitem{Giombi:2011ya}
S.~Giombi and X.~Yin, ``{On Higher Spin Gauge Theory and the Critical O(N)
  Model},'' {\em Phys. Rev.} {\bf D85} (2012) 086005,
  \href{http://xxx.lanl.gov/abs/1105.4011}{{\tt 1105.4011}}.

\end{thebibliography}
\end{document}